\documentclass[12pt,a4paper]{article}
\usepackage{jheplike,tocloft,ifpdf,amsmath,mathrsfs,xcolor,cancel}

\widowpenalty=500\clubpenalty=1000\unitlength=1mm\textheight=8.7in\hypersetup{pdftitle={},pdfcreator={},linkcolor=[rgb]{0.15,0.35,0.75},colorlinks=true,citecolor=[rgb]{0.675,0,0.2},urlcolor=[rgb]{0.15,0.35,0.65}}
\makeatletter\renewcommand\section{\addtocontents{toc}{\protect\addvspace{-2.25\p@}}\@startsection {section}{1}{\z@}{0.5ex \@plus .2ex \@minus 0.2ex}{0.3ex \@plus.1ex\@minus .5ex}{\normalfont\large\bfseries}}
\renewcommand\subsection{\addtocontents{toc}{\protect\addvspace{0.5\p@}}\@startsection {subsection}{1}{\z@}{0.5ex \@plus .2ex \@minus 0.2ex}{0.3ex \@plus.1ex\@minus .5ex}{\normalfont\bfseries}}
\renewcommand\subsubsection{\addtocontents{toc}{\protect\addvspace{-2.5\p@}}\@startsection {subsubsection}{1}{\z@}{0.5ex \@plus .2ex \@minus 0.2ex}{0.3ex \@plus.1ex\@minus .5ex}{\normalfont\bfseries}}
\newcommand{\eq}[1]{\vspace{-0.5pt}\begin{equation}#1\vspace{-0.5pt}\end{equation}}
\newcommand{\fwbox}[2]{\text{\makebox[#1][c]{$\hspace{-150pt}\displaystyle#2\hspace{-150pt}$}}}
\newcommand{\fwboxL}[2]{\text{\makebox[#1][l]{$#2$}}}
\newcommand{\fwboxR}[2]{\text{\makebox[#1][r]{$#2$}}}
\newcommand{\equivR}{\fwbox{15.5pt}{\hspace{-0pt}\fwboxR{0pt}{\raisebox{0.47pt}{\hspace{1.25pt}:\hspace{-4pt}}}=\fwboxL{-0pt}{}}}
\newcommand{\equivL}{\fwbox{13.5pt}{\fwboxR{-2pt}{}=\fwboxL{0pt}{\raisebox{0.47pt}{\hspace{-4pt}:\hspace{1.25pt}}}}}
\newcommand{\fig}[3]{\raisebox{#1}{\includegraphics[scale=#2]{#3}}}
\newcommand{\bigger}[1]{\raisebox{-0.95pt}{\scalebox{1.25}{$#1$}}}
\newcommand{\x}[2]{{\color{black}(}\hspace{-0.85pt}{\color{black}#1}\hspace{-0.25pt}{\color{black}|}\hspace{-0.25pt}{\color{black}#2}\hspace{-0.85pt}{\color{black})}}
\newcommand{\embd}[1]{{\color{black}|}\hspace{-0.5pt}{\color{black}#1}\hspace{-0.75pt}{\color{black})}}
\newcommand{\smrt}[1]{\sqrt{\phantom{\fwbox{0pt}{b}}\smash{#1}}}
\newcommand{\proj}[1]{\left[\!#1\!\,\rule{0pt}{10pt}\right]}
\newcommand{\gram}{\mathcal{G}}
\newcommand{\ru}[2]{r^{[#1]}_{\!#2}}
\newcommand{\rs}[1]{\bar{r}_{\!#1}}

\DeclareMathOperator*{\Res}{\mathrm{Res}}
\DeclareMathOperator{\arccosh}{arccosh}
\DeclareMathOperator{\li}{Li}
\renewcommand{\phi}{\varphi}
\renewcommand{\bar}{\overline}
\renewcommand{\hat}{\widehat}

\definecolor{hred}{rgb}{0.75,0.0,0.325}
\definecolor{hgreen}{rgb}{0.2,0.605,0}
\definecolor{hblue}{rgb}{0,0,0.975}

\title{\texorpdfstring{~\\[0pt]{\LARGE \mbox{All-Mass $n$-gon Integrals in $n$ Dimensions}}\\[-24pt]}{All-Mass $n$-gon Integrals in $n$ Dimensions}}
\author[a,b,c]{\vspace{-24pt}Jacob~L.~Bourjaily,}\emailAdd{bourjaily@psu.edu}
\author[d]{Einan~Gardi,}\emailAdd{einan.gardi@ed.ac.uk}
\author[a]{Andrew~J.~McLeod,}\emailAdd{amcleod@nbi.ku.dk}
\author[a]{Cristian~Vergu}\emailAdd{c.vergu@gmail.com}
\affiliation[a]{Niels Bohr International Academy and Discovery Center, Niels Bohr Institute,\\University of Copenhagen, Blegdamsvej 17, DK-2100, Copenhagen \O, Denmark}
\affiliation[b]{Center for the Fundamental Laws of Nature, Department of Physics,\\ Jefferson Physical Laboratory, Harvard University, Cambridge, MA 02138, USA}
\affiliation[c]{Institute for Gravitation and the Cosmos, Department of Physics,\\Pennsylvania State University, University Park, PA 16892, USA}
\affiliation[d]{Higgs Centre for Theoretical Physics, School of Physics and Astronomy,\\The University of Edinburgh, Edinburgh EH9 3FD, Scotland, UK}

\abstract{%
We explore the correspondence between one-loop Feynman integrals and (hyperbolic) simplicial geometry to describe the \emph{all-mass} case: integrals with generic external and internal masses. Specifically, we focus on $n$-particle integrals in exactly $n$ space-time dimensions, as these integrals have particularly nice geometric properties and respect a dual conformal symmetry. In four dimensions, we leverage this geometric connection to give a concise dilogarithmic expression for the all-mass box in terms of the Murakami-Yano formula. In five dimensions, we use a generalized Gauss-Bonnet theorem to derive a similar dilogarithmic expression for the all-mass pentagon. We also use the Schl\"afli formula to write down the symbol of these integrals for all $n$. Finally, we discuss how the geometry behind these formulas depends on space-time signature, and we gather together many results related to these integrals from the mathematics and physics literature.
}
\preprint{}

\begin{document}
\maketitle\thispagestyle{empty}

%================================================================================================================
%    1. Introduction 
%================================================================================================================
%\newpage
\setcounter{page}{1}\vspace{-0pt}%
\pagenumbering{roman}\clearpage\pagenumbering{arabic}
\vspace{-6pt}\section{Introduction and and Overview}\label{sec:introduction}\vspace{-0pt}
%================================================================================================================

Among the broad class of special functions that emerge in our description of scattering amplitudes in perturbative quantum field theory, polylogarithms play a special role. Not only are these functions under the best theoretical control,  they also prove sufficient to describe one-loop scattering processes (in any theory, for any number of dimensions). This ubiquity follows from integral reduction combined with the fact that any one-loop Feynman integral (in any integer number of dimensions) can be expressed in terms of generalized polylogarithms. Although more complicated transcendental functions are known to appear in generic scattering processes at higher loop orders, polylogarithms also prove sufficient to describe many low-multiplicity processes beyond one loop (and sometimes, perhaps, to all loop orders). 

In this paper, we study the class of polylogarithms that appear as one-loop Feynman integrals in generic quantum field theories. In particular, we are interested in the most general (or universal) form of these integrals, corresponding to the case in which all external and internal masses are taken to be generic. We call these \emph{all-mass} integrals. We focus here on $n$-particle integrals in exactly $n$ space-time dimensions, which prove to have particularly nice geometric properties and respect a dual conformal symmetry. In a companion paper,~\cite{n_gons_d_dim}, we will explore a similar set of ideas for the case of all-mass $n$-particle integrals in a generic number of space-time dimensions. Dimensional shift identities~\cite{Bern:1992em,Tarasov:1996br,Lee:2009dh} can also be used to relate the functions we study here to integrals in other integer dimensions.

These \(n\)-gon integrals constitute a physically interesting and instructive class of examples for developing our understanding quantum field theory. They are sufficiently complex to exhibit many of the expected features of higher-loop Feynman integrals, yet are already understood from a diverse set of geometric and computational perspectives. In particular, these integrals have a geometrical interpretation as volumes of geodesic simplices in hyperbolic space (as studied in~\cite{Davydychev:1997wa,Mason:2010pg}), making it possible to leverage powerful techniques from the mathematics literature for their computation.

The study of these integrals has a long history.  In particular, the box integral has been studied in the physics literature by Wu~\cite{actwu:1961}, \mbox{'t Hooft} and Veltman~\cite{tHooft:1978jhc}, Denner, Nierste, and Scharf~\cite{Denner:1991qq}, and Hodges \cite{Hodges:2010kq}.  The pentagon integral in five dimensions with massless propagators has also been studied by Nandan, Paulos, Spradlin, and Volovich~\cite{Nandan:2013ip}.  Earlier mathematical studies include~\cite{aomoto1977, MR1239859, MR1338325, MR1649192}, and results for $n$-gon Feynman integrals can be found in~\cite{Ellis:2007qk,Dixon:2011ng,DelDuca:2011ne,DelDuca:2011jm,DelDuca:2011wh,Papadopoulos:2014lla,Spradlin:2011wp,Kozlov:2015kol}. In particular, previous papers that have made use of the correspondence between one-loop Feynman integrals and hyperbolic volumes include~\mbox{\cite{aomoto1977,Aomoto1992,MR1649192,Davydychev:1997wa,Mason:2010pg,Schnetz:2010pd,Spradlin:2011wp,Nandan:2013ip,Davydychev:2017bbl,Arkani-Hamed:2017ahv,Herrmann:2019upk}}.  Recently, an approach based on Yangian symmetry has also been discussed~\cite{Loebbert:2019vcj}.

We build on this literature by first presenting new formulas for the all-mass box in four dimensions, making use of the Murakami-Yano formula for the volume of a hyperbolic tetrahedron~\cite{MR2154824}, as well as a similar formula for the volume of a tetrahedron in spherical (or Euclidean) signature~\cite{MR2917101}. An interesting feature of these formulas is that they depend on the angles formed at the vertices of these simplices, rather than on the lengths of their edges; as a result, they take an especially parsimonious dilogarithmic form. Using these formulas, we write down concise expressions for the all-mass box integral that make its permutation and conformal symmetries manifest, and which only involve a single algebraic root. We also derive an expression that is valid in all (four-dimensional) space-time signatures, whose arguments are more directly related to the external kinematics of the Feynman integral.

While explicit results for the all-mass box have long existed in the literature~\mbox{\cite{actwu:1961,tHooft:1978jhc,Denner:1991qq,Hodges:2010kq}}, one-loop integrals provide an ideal laboratory in which to explore the most natural functions and variables for expressing (the transcendental part of) higher-loop integrals. As such, we deem it worthwhile to work towards increasingly compact and elegant expressions for integrals that promise to be instructive in this regard---a criteria that the all-mass box, which famously involves algebraic roots, certainly satisfies. In particular, we consider the formulas presented here to have significant advantages over previous ones presented in the literature with respect to symmetries, domains of validity, and simplicity.

Building on these results, we also derive an explicit formula for the all-mass pentagon integral in five dimensions using a generalized Gauss-Bonnet theorem (see~\cite{Schnetz:2010pd}). These results, valid in hyperbolic and spherical signature, again manifest the permutation and conformal invariance of these integrals, and involve just a five-orbit of algebraic roots.

Using the correspondence with simplicial volumes, the symbol~\cite{Goncharov:2010jf} of these integrals can also be computed for any number of particles using the Schl\"afli formula~\cite{Schlaefli:1860}. We give explicit formulas for these symbols that are valid for all $n$. Notably, this class of integrals includes members of arbitrarily high transcendental weight, as the weight of these integrals grows linearly with particle multiplicity. Similar results for one-loop symbols can be found in~\cite{MR1649192,Spradlin:2011wp,Arkani-Hamed:2017ahv,Abreu:2017mtm}. In particular, we find a marked correspondence with the results of~\cite{Abreu:2017mtm}, which were derived using different (motivic) methods, and which arise from a different, more graph-theoretic, perspective on Feynman integrals. 

Although in this work we carry out only a cursory investigation of the (all-$n$) analytic structure of these integrals, it is our hope that this class of symbols will prove useful for developing our understanding of the (more general) analytic properties of Feynman integrals, and especially for developing methods by which symbol alphabets can be (predictively) tailored to individual Feynman diagrams and amplitudes (see also~\cite{Bourjaily:2018aeq} for some work in this direction). \\

The organization of the paper is as follows. We first define the class of integrals under study and discuss their normalization, which can be chosen to yield unit leading singularities. These integrals can be expressed in terms of dual variables, and are invariant under a (dual) conformal symmetry. In section~\ref{sec:hyperbolic_geometry}, we review various aspects of hyperbolic geometry, and then show how an exact correspondence can be made between the volumes of hyperbolic simplices and $n$-gon Feynman integrals in $n$ dimensions with the choice of a reference point at infinity ~\cite{Hodges:2010kq}. We also discuss how similar correspondences hold with simplices in different signatures outside of Lorentzian kinematics. In section~\ref{sec:loop_integrals}, we work out examples of this correspondence in low dimensions, studying the bubble integral in two dimensions and the triangle integral in three dimensions.  Then, in section~\ref{subsec:box}, we make use of known volume formulas for tetrahedra in hyperbolic and spherical signatures (from Murakami and Yano) to give new formulas for the all-mass box integral. In this section we also derive a formula that works in all space-time signatures, and study how these formulas simplify in a dual conformal light-like limit. In section~\ref{sec:odd_n_integrals}, we present a discussion of the Gauss-Bonnet theorem for manifolds with corners, which can be applied to compute the volume of \(n\)-dimensional simplices in terms of $(n{-}1)$-dimensional simplices when \(n\) is odd.  Using this method, we obtain explicit formulas for the all-mass pentagon integral in five dimensions in both hyperbolic and spherical signatures. We additionally show how these results simplify when one or more of the internal masses goes to zero.  Finally, in section~\ref{sec:schlafli}, we use the Schl\"afli formula to derive an explicit formula for the symbol of these integrals for any $n$, and study certain aspects of their branch cut structure.  We end with some conclusions, and by outlining some open questions.

We also include a short introduction to the embedding formalism in appendix~\ref{appendix:embedding_formalism}, as it is from this perspective that the dual conformal invariance of these integrals is most readily seen.

%================================================================================================================
\vspace{-0pt}\subsection{All-Mass \texorpdfstring{$n$}{n}-gon Feynman Integrals in \texorpdfstring{$n$}{n} Dimensions}\label{subsec:all_mass_integrals_basic_definitions_and_notation}\vspace{-0pt}
%================================================================================================================

\begin{figure}[t]\centering\vspace{-10pt}$$\fig{-50pt}{1}{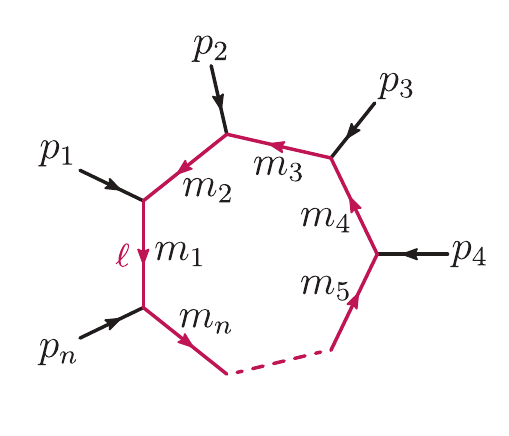}\bigger{\Leftrightarrow}\fig{-50pt}{1}{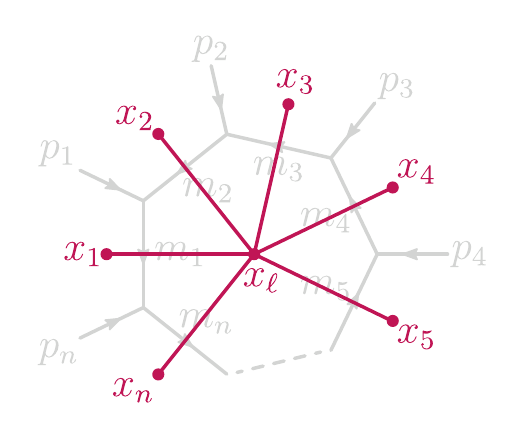}\vspace{-30pt}$$\caption{The $n$-point, all-mass integral and its dual-momentum space representation.}\label{fig:integral_and_dual}
\end{figure}

We are interested in the scalar Feynman integral shown in \mbox{Figure \ref{fig:integral_and_dual}}, where the loop momentum $\ell$ is $n$-dimensional, and all the external momenta and internal masses are taken to be generic: $p_i^2\neq0$, $m_i\neq 0$. We may define this integral in (all-plus) Euclidean-signature to be\footnote{In momentum space, the loop integration measure should also include a factor of $1/(2\pi)^n$. We leave this off because it would be scaled-out anyway soon---as explained in the next footnote.}
\eq{I_n^0\equivR\int\!\!\!d^n\ell\,\frac{1}{\big[\ell^2+m_1^2\big]\big[(\ell-p_1)^2+m_2^2\big]\cdots\big[(\ell-(p_1+\cdots+p_{n-1}))^2+m_n^2\big]}\,.\label{integral_definition_in_space-time}}
(We will have more to say about other space-time signatures in section~\ref{subsec:space-time_signatures}.) Notice that we have decorated $I_n^0$ with a superscript `$0$' to emphasize that we will soon have reason to change its normalization. 

In order to manifest momentum conservation and the invariance of (\ref{integral_definition_in_space-time}) under translations of the loop momentum $\ell$, we introduce {\it dual-momentum} coordinates $\{x_i\}$ such that \mbox{$p_i\equivL(x_{i+1}-x_i)$}, with cyclic indexing understood. In terms of these coordinates, it is easy to see that consecutive sums of external momenta appearing in the propagators of (\ref{integral_definition_in_space-time}) become squared differences:
\begin{align}
I_n^0&=\int\!\!\!d^n\ell\,\frac{1}{\big[(\ell-(x_1-x_1))^2+m_1^2\big]\big[(\ell-(x_2-x_1))^2+m_2^2\big]\cdots\big[(\ell-(x_n-x_1)^2+m_n^2\big]}\nonumber\\
&\equivL\int\!\!\!d^n\!x_{\ell}\,\frac{1}{\big[(x_\ell-x_1)^2+m_1^2\big]\big[(x_{\ell}-x_2)^2+m_2^2\big]\cdots\big[(x_{\ell}-x_n)^2+m_n^2\big]}\nonumber\\
&\equivL\int\!\!\!d^n\!x_{\ell}\,\frac{1}{\big(x_{\ell\,1}^2+m_1^2\big)\big(x_{\ell\,2}^2+m_2^2\big)\cdots\big(x_{\ell\,n}^2+m_n^2\big)}\,,\label{dual_momentum_space_definition}
\end{align}
where in the second step we defined the dual loop-momentum variable $x_{\ell}$ according to \mbox{$\ell\equivL x_{\ell}-x_1$} and in the last step we introduced the familiar notation for dual-momentum Mandelstam invariants, $x_{ij}^2\equivR(x_j-x_i)^2$. 

Introducing Feynman parameters in the canonical way (and doing the standard translations and rescalings), it is not hard to express (\ref{dual_momentum_space_definition}) as
\eq{\begin{split}I_n^0=\Gamma(n)\int\limits_0^\infty\!\!\!\proj{d^{n-1}\!\vec{\alpha}}\int\!\!\!d^n\!x_\ell\frac{1}{\big[x_{\ell}^2+\mathscr{F}\big]^n}=\pi^{n/2}\Gamma(n/2)\int\limits_0^\infty\!\!\!\proj{d^{n-1}\!\vec{\alpha}}\frac{1}{\mathscr{F}^{\frac{n}{2}}}\end{split}\,,\label{all_mass_param_integral_0}}
where $\mathscr{F}$ is the second Symanzik polynomial
\eq{\mathscr{F}\equivR\Big[\sum_{i}\alpha_i^2m_i^2\Big]+\sum_{i<j}\alpha_i\alpha_j\big(x_{ij}^2+m_i^2+m_j^2\big)\,\label{symanzik_f_defined}}
and we have used $\proj{d^{n-1}\!\vec{\alpha}\rule{0pt}{10pt}}$ to denote the canonical volume form on the projective space $\mathbb{RP}^{n-1}$ of Feynman parameters
\eq{\proj{d^{n-1}\!\vec{\alpha}\rule{0pt}{10pt}}\equivR\sum_{i = 1}^n (-1)^i \alpha_i \, d \alpha_1 \wedge \cdots \wedge \widehat{d \alpha_i} \wedge \cdots \wedge d \alpha_n\,.\label{projective_notation}}
This volume form is frequently written with an explicit choice of de-projectivization\\[-8pt]
\eq{\proj{d^{n-1}\!\vec{\alpha}}\simeq d^n\!\vec{\alpha}\,\,\,\delta(\alpha_i-1)\label{projective_measure_with_explicit_deprojectivization}}
for any choice of $\alpha_i$. Notice that Feynman's preferred choice of de-projectivization, $\delta\big(\sum_i\alpha_i-1\big)$, is related to that of (\ref{projective_measure_with_explicit_deprojectivization}) by a change of variables with unit Jacobian. 

It will be useful to re-express the second Symanzik polynomial (\ref{symanzik_f_defined}) in a somewhat more compact way. In particular, we introduce an $n\times n$ matrix $\gram^0$ with components
\eq{\gram^0_{ij}\equivR\frac{1}{2}\big(x_{ij}^2+m_i^2+m_j^2\big)\label{defn_of_gram0}}
so that 
\eq{\mathscr{F}=\sum_{i,j}\gram^0_{ij}\alpha_i\alpha_j\,.\label{symanzik_f_in_terms_of_gram0}}
The factor of $\frac{1}{2}$ in (\ref{defn_of_gram0}) is a symmetry factor, allowing us to write (\ref{symanzik_f_in_terms_of_gram0}) more obviously as matrix multiplication: $\mathscr{F}=\vec{\alpha}^T\!\!.\gram.\vec{\alpha}$ where $\vec{\alpha}\equivR\!(\alpha_1,\ldots,\alpha_n)$.

\paragraph{Leading Singularities and Purity}~\\[-12pt]

$I_n^0$ as defined in (\ref{dual_momentum_space_definition}) is an $n$-dimensional integral with $n$ loop-dependent factors in its denominator. Importantly, it has {\it leading singularities}: residues of maximal co-dimension. It is canonical to normalize such integrals so that (at least some choice of) leading singularities are unit in magnitude. An integral with the property that all its leading singularities are unit in magnitude is called {\it pure} \cite{ArkaniHamed:2010gh}. The integral $I_n^0$ is known to be pure up to a constant of normalization---fixed by any one of its leading singularities.

The calculation of the maximal co-dimension residues of $I_n^0$ is not entirely trivial (although it is significantly easier in the embedding space formalism discussed in appendix~\ref{appendix:embedding_formalism}); therefore, we merely quote the fact that there are always {\it two} leading singularities which cut all $n$ propagators, and that these leading singularities are
\eq{\Res_{\left\{x_{\ell\,i}^2+m_i^2=0\right\}}\!\!\Bigg(\!\frac{d^n\!x_{\ell}}{\big(x_{\ell\,1}^2+m_1^2\big)\big(x_{\ell\,2}^2+m_2^2\big)\cdots\big(x_{\ell\,n}^2+m_n^2\big)}\!\Bigg)=\frac{\pm1\phantom{\pm}}{2^n\sqrt{\det\gram^0}}\,.\label{leading_singularties}}
Because of this, 
\eq{I_n\equivR2^n\sqrt{\det\gram^0}I_n^0\,\label{pure_integral_defined}}
will have `unit leading singularities' and is in fact pure. 

Notice that, although the integral $I_n^0$ is positive definite (on the principal branch) for real kinematics, $I_n$ may not be: for example, when $\det\gram^0<0$, our definition of $I_n$ will be pure imaginary. This is a convention; we could have chosen instead to use $\sqrt{|\det\gram^0|}$ in the normalization of (\ref{pure_integral_defined}), but the choice we have made is the more standard one (and the one we find will allow for slightly simpler formulas below).  As we will see, however, it will be useful to sometimes make use of  
\eq{\sigma(\gram^0)\equivR\mathrm{sign}(\det\gram^0\!)\,.\label{sign_of_signature}}
With this normalization,\footnote{Notice that the factor of $1/(2\pi)^n$ `missing' from (\ref{integral_definition_in_space-time}) would have also appeared in (\ref{leading_singularties}) then dropped out of the definition of $I_n$ in (\ref{pure_integral_defined}).} the Feynman integral (\ref{all_mass_param_integral_0}) becomes
\eq{I_n=(2\sqrt{\pi})^n\Gamma(n/2)\int\limits_0^\infty\!\!\!\proj{d^{n-1}\!\vec{\alpha}}\frac{\sqrt{\det\gram^0}}{\big(\sum_{ij}\gram^0_{ij}\alpha_i\alpha_j\big)^{\frac{n}{2}}}\,,\label{conformal_int_penult}}
where we have adopted the notation in~\eqref{symanzik_f_in_terms_of_gram0}.

In addition to being pure, the integral $I_n$ is known to have transcendental weight $n$~\cite{Arkani-Hamed:2017ahv}. Isolating the kinematic-dependent integral as $\hat{I}_n$ via
\eq{\hat{I}_n \equivR \frac{1}{(2\sqrt{\pi})^n\Gamma(n/2)}I_n\,,\label{I_hat}}
we cleanly separate this weight into two parts: the prefactor we have divided out has transcendental weight $\lceil n/2\rceil$, while the integral $\hat{I}_n$ has weight $\lfloor n/2\rfloor$.

\paragraph{Something a Little \emph{Odd} About the `Scalar' Integral $I_n$}~\\[-12pt]

The original integral $I_n^0$ (\ref{integral_definition_in_space-time}) was built from ordinary scalar Feynman propagators. Its overall sign (or phase) is intrinsically well defined, including its dependence on space-time signature. In contrast, the pure integral $I_n$ defined by (\ref{pure_integral_defined}) has a \emph{conventional} overall sign. Even fixing branch conventions for $\sqrt{\det\gram}$, multidimensional residues are intrinsically \emph{oriented} quantities whose signs depend on the orientation of the contour integral (or the ordering of integration variables in the Jacobian) that defines them. 

Because the left hand side of (\ref{leading_singularties}) should be viewed as oriented---antisymmetric in the ordering of the propagators, say---we might choose to view the normalization of $I_n$ in (\ref{pure_integral_defined}) as also carrying this orientation thereby rendering $I_n$ anti-cyclic in even-dimensional spaces. This corresponds to interpreting (\ref{conformal_int_penult}) as an \emph{oriented} integral. We do not take this view here, mostly for practical (and for notational) reasons. However, we emphasize that the sign of the normalized integral $I_n$ corresponds to a choice of convention. 

\newpage 
\paragraph{Scale Invariance and Conformality}~\\[-12pt]

The integral $I_n$ would seem to depend on $\binom{n}{2}$ Mandelstam invariants $x_{ij}^2$ and $n$ internal masses. However, this integral has a hidden conformal symmetry. To see this, we first re-write (\ref{conformal_int_penult}) to remove the dimensionful parameters in the matrix $\gram^0$. One way to do this is to rescale the Feynman parameters according to\footnote{The reader should forgive our abuse of notation in using $\alpha_i$ to denote the integration variable both before and after the rescaling (\ref{alpha_rescaling}).}
\eq{\alpha_i\mapsto\alpha_i/m_i\,.\label{alpha_rescaling}}
This introduces a Jacobian of $1/(\prod_im_i)$, resulting in
\begin{align}I_n\underset{\text{(\ref{alpha_rescaling})}}{\longmapsto}&(2\sqrt{\pi})^n\Gamma(n/2)\int\limits_0^\infty\!\frac{\proj{d^{n-1}\!\vec{\alpha}}}{\prod_im_i}\frac{\sqrt{\det\gram^0}}{\Big(\sum_{ij}\big(\gram^0_{ij}/(m_im_j)\big)\alpha_i\alpha_j\Big)^{\frac{n}{2}}}\nonumber\\
=&(2\sqrt{\pi})^n\Gamma(n/2)\int\limits_0^\infty\!\!\!\proj{d^{n-1}\!\vec{\alpha}}\frac{\sqrt{\det\gram}}{\big(\sum_{ij}\gram_{ij}\alpha_i\alpha_j\big)^{\frac{n}{2}}}\,,\label{conformal_int_ult}
\end{align}
where we have introduced a new matrix $\gram$ that has entries
\eq{\gram_{ij}\equivR\gram^0_{ij}/(m_im_j)=\frac{x_{ij}^2+m_i^2+m_j^2}{2\,m_i\,m_j}\,.\label{gram_matrix}}
Note that $\gram$ is symmetric and has $1$ in its diagonal entries, so it depends on just $n(n-1)/2$ independent pieces of kinematic data. We can think of $I_n(\gram)$ as being a function directly of this matrix $\gram$.

Not only is it clear now that $I_n(\gram)$ is scale-invariant (under a simultaneous transformation of all $(x_a^\mu,m_a)\mapsto(\lambda\,x_a^\mu,\lambda\,m_a)$), but it turns out to also be fully conformally invariant. This fact is hinted at by the structural equivalence between (\ref{conformal_int_penult}) and (\ref{conformal_int_ult}), and can be made concrete by noting the invariance of $I_n$ under the inversion
\eq{x_i^\mu \to \frac {x_i^\mu} {x_i^2 + m_i^2} , \qquad m_i \to \frac {m_i} {x_i^2 + m_i^2} , \qquad x_{\ell}^\mu \to \frac {x_{\ell}^\mu} {x_{\ell}^2}\,.}
This conformal invariance can be better understood from the viewpoint of the embedding formalism, which we discuss in more detail in appendix~\ref{appendix:embedding_formalism}.
%

%================================================================================================================
%    2. Hyperbolic Geometry Section
%================================================================================================================
\newpage\vspace{-0pt}\section[Hyperbolic Geometry and Kinematic Domains]{\mbox{\hspace{-3pt}Hyperbolic Geometry and Kinematic Domains}}\label{sec:hyperbolic_geometry}\vspace{-0pt}
%================================================================================================================

Let us now turn to the computation of volumes in hyperbolic space. We start by considering the space \(\mathbb{E}^{n-1,1}\), which we take to be $n$-dimensional Euclidean space equipped with the Lorentzian scalar product 
\eq{\langle x, y\rangle \equivR x_1 y_1 + \cdots + x_{n-1} y_{n-1} -  x_n y_n \, \label{eq:ambient_metric}}
for any vectors $x, y \in \mathbb{E}^{n-1,1}$. In this space we distinguish three types of vectors: those that are `time-like' (\(\langle x, x\rangle < 0\)); those that are `light-like' (\(\langle x, x\rangle = 0\)); and those that are space-like (\(\langle x, x\rangle > 0\)).  In the case of time-like and light-like vectors, we further differentiate vectors whose last component is positive or negative.

The collection of time-like vectors that satisfy \(\langle x, x\rangle = -1\) and \(x_n > 0\) define one branch of a hyperboloid (which we will refer to as its positive branch).  This space of vectors furnishes one realization of hyperbolic space \(\mathbb{H}^{n-1}\) and constitutes the \emph{hyperboloid model}.  Making the change of variables \(x_n = \cosh \tau\) and \(x_i = z_i \sinh\tau\) for \mbox{\(i = 1, \dots, n-1\)}, this hyperboloid constraint becomes the requirement that the $z_i$ lie on the unit sphere: \mbox{\(z_1^2 + \cdots+ z_{n-1}^2 = 1\)}.  It follows that the inner product (\ref{eq:ambient_metric}) induces the metric
\eq{d s^2\!= d \tau^2 + \sinh^2\!\tau \,\,d \Omega_{n - 2}^2\,,}
where \(d \Omega_{n - 2}^2\) is the measure on the (\(n{-}2\))-dimensional unit sphere.  Hence, the induced metric from the embedding space is a Riemannian metric.

Starting from any two points \(x, y\) on the positive branch of this hyperboloid, we can rotate our coordinate system on \(\mathbb{E}^{n-1,1}\) so that we have \mbox{\(x = (0,\dots, 0, 1)\)} and \mbox{\(y = (0, \dots, 0, \sinh\tau, \cosh\tau)\)}.  The geodesic curve through \(x\) and \(y\) is given by \(( 0, \dots, 0, \sinh t, \cosh t)\) for $0 \leq t \leq \tau$, and the line element along this geodesic is $d s^2 = d \tau^2$.  Since $\langle x,y\rangle=-\cosh\tau$, the hyperbolic distance \(d(x, y)\) between \(x\) and~\(y\) along the geodesic that joins them is
\begin{equation} \label{eq:hyperbolic_distance}
  d(x, y) \equivR \tau = \arccosh \big(-\langle x, y\rangle\big).
\end{equation}
Similarly, it is easy to see that the volume form \(d x_1 \cdots d x_n\) in \(\mathbb{E}^{n-1,1}\) induces the form
\begin{equation} \label{eq:vol_hyperboloid}
  d\text{vol} \equivR \delta \big(\langle x, x\rangle + 1\big) \theta(x_n)d x_1 \cdots d x_n =
  \frac {\delta \big(x_n - \smrt{1 + x_1^2 + \cdots +x_{n-1}^2} \big)}{2 \smrt{1 + x_{1}^2 + \cdots+ x_{n-1}^2}} \theta(x_n)d x_1 \cdots d x_n
\end{equation}
on the upper branch of the hyperboloid.

\begin{figure}[t]\centering\fig{-10pt}{1}{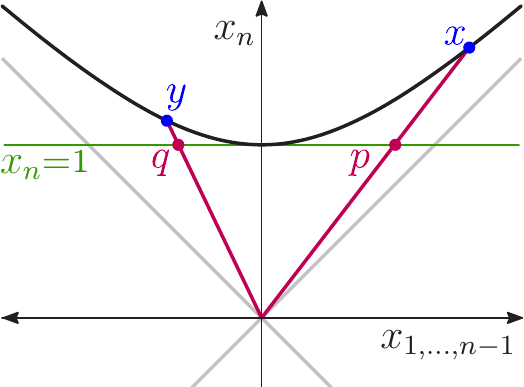}
\caption{The hyperboloid and projective models of hyperbolic space, as they appear embedded in \(\mathbb{E}^{n-1,1}\).  In the hyperboloid model, points in hyperbolic space belong to the upper branch of the hyperboloid, while in the projective model they belong to the \(x_n = 1\) hyperplane.  The points in these two models are in one-to-one correspondence, and are identified when they lie on the same line passing through the origin of the embedding space.}\label{fig:hyperbolic_models}
\end{figure}

There are several other ways to represent hyperbolic space. Another representation that will prove useful for us is the \emph{projective model} (sometimes called the Klein model).  This model realizes hyperbolic space as the set of lines that intersect both the origin and the upper branch of the hyperboloid considered above, as show in Figure~\ref{fig:hyperbolic_models}.  Some of these lines are tangent to the upper branch of the hyperboloid; these lines correspond to the boundary of hyperbolic space. While geodesic lines and hypersurfaces correspond to straight lines and planes in the projective model, it breaks the conformal symmetry insofar as rotations of the original embedding space \(\mathbb{E}^{n-1,1}\) do not preserve angles.

For every point $\smash{x = \big(x_1, \dots, x_{n-1},\sqrt{\phantom{\fwbox{0pt}{b}}\smash{1\!+ x_1^2 + \cdots + x_{n-1}^2}} \big)}$ in the upper branch of the hyperboloid, the corresponding point in the projective model is given by $p =$ $(p_1, \dots, p_{n-1},1)$, where \(p_i\equivR x_i / \sqrt{\phantom{\fwbox{0pt}{b}}\smash{1 + x_1^2 + \cdots + x_{n-1}^2}}\); equivalently, we could view $x_i \equivR$ $p_i / \smrt{\phantom{\fwbox{0pt}{b}}\smash{1 - p_1^2 - \cdots - p_{n-1}^2}}$. This maps the upper branch of the hyperboloid to the interior of the unit ball in the plane \(x_n = 1\), centered at \((0, \dots, 0,1) \in \mathbb{E}^{n-1,1}\). We denote the inner product of two points $p$ and $q$ in the projective model by
\begin{align} \label{eq:projective_metric}
Q(p, q) = 1 - \sum_{i=1}^{n-1} p_i q_i \, .
\end{align}  
Note that the metric $Q(p, q)$ differs from the metric of the ambient space by a non-constant rescaling $p_i\mapsto p_i/\sqrt{Q(p_i,p_i)}$, which maps the points at infinity to the boundary of the unit ball defined by $Q(p, p) = 0$. In these coordinates,~\eqref{eq:vol_hyperboloid} becomes
\begin{equation} \label{eq:projective_model_volume_form}
  d\text{vol} = \frac 1 2 \frac {d p_1 \cdots d p_{n-1}}{Q(p,p)^{\frac n 2}} \, ,
\end{equation}
where now $p_i^2 \leq 1$.

\begin{figure}[t]\centering\fig{-10pt}{1}{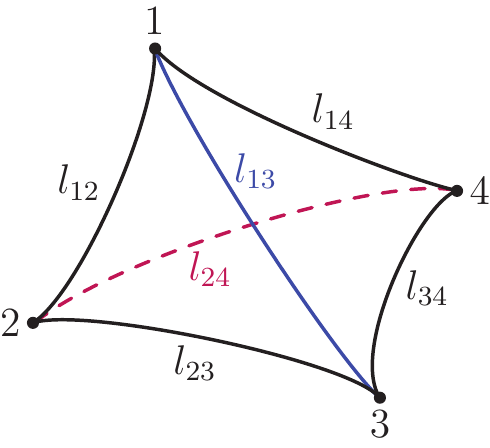}\caption{A three-dimensional simplex (a tetrahedron) in hyperbolic space $\mathbb{H}^3$.}\label{fig:tetrahedron}
\end{figure}

Now consider an (\(n{-}1\))-simplex with vertices \(v_1, \dots, v_n \in \mathbb{E}^{n-1,1}\) such that the last component of each \(v_i\) is equal to unity.\footnote{Thus far, we have used indices on $x$ and $y$ (in the hyperboloid model) and $p$ and $q$ (in the projective model) to denote components. We now switch to a notation where indices on $v$ (in the projective model) and $h$ (in the hyperboloid model, in the next section) denote distinct points.} The interior points of this simplex can be parametrized by
\begin{equation} \label{eq:n-simplex}
  p(\beta) = \sum_{i = 1}^n \beta_i v_i,
\end{equation}
where \(\beta_i > 0\) and \(\sum_{i = 1}^n \beta_i = 1\). Using the \(\beta_i\) variables, the numerator of~\eqref{eq:projective_model_volume_form} can be rewritten as
\begin{align}
  d p_1(\beta) \cdots d p_{n-1}(\beta) &= \
  \det_{i, j} (v_i - v_n)_j \  d \beta_1 \cdots d \beta_{n-1} \\
  &= 
  \vert v_1 \wedge \cdots \wedge v_n\vert \ d \beta_1 \cdots d \beta_{n-1}\,.
\end{align}
Furthermore, we have 
\begin{equation}
  |\det\!\big(Q_{ij}\big)| =  \vert v_1 \wedge \cdots \wedge v_n\vert^2\equivR\det\!\big(v_1,\ldots,v_n\big)^2\,,
\end{equation}
where we have defined $Q_{ij}$ as the matrix with entries $Q_{i,j}\equivR Q(v_i,v_j)$.  
Putting these results together, \eqref{eq:projective_model_volume_form} can be rewritten as
\begin{equation} \label{eq:beta_variables_volume}
  d\text{vol}\left(Q(v_i, v_j) \right) = \frac 1 2 \frac {\sqrt{|\det(Q_{i,j})|} \, d \beta_1 \cdots d \beta_{n-1}}{\left(\sum_{i, j} Q_{ij} \beta_i \beta_j \right)^{\frac n 2}}\,.
\end{equation}
Finally, we make a change of variable \(\beta_i\mapsto\alpha_i/(\sum_{i}\alpha_i)\) to obtain
\begin{align}
  \label{eq:projective-volume}
  d\text{vol}\left(Q(v_i, v_j) \right) = \frac 1 2 \proj{d^{n-1}\!\vec{\alpha}} \frac {\sqrt{\left|\det(Q_{i,j})\right|}}{\left(\sum_{i, j} Q_{ij} \alpha_i \alpha_j \right)^{\frac n 2}}\,,
\end{align}
where \(0 < \alpha_i < \infty\) and (since $\alpha_n \neq 1$) we have lifted the differential form in~\eqref{eq:beta_variables_volume} to the full projective measure~\eqref{projective_notation}. 

Let us now pause to highlight the fact that the volume~\eqref{eq:projective-volume} is precisely the one-loop $n$-point Feynman integral given in~\eqref{conformal_int_ult}, up to some numerical prefactor and the fact that the latter integral has been de-projectivized by the choice $\alpha_n=1$. The points of the simplex whose volume we are calculating are encoded by kinematics via the matrix $\gram$. 

Before exploring the connections between kinematics and the geometry of hyperbolic simplices, we note that the cases of even and odd \(n\) are qualitatively different.  When \(n\) is even the volume form is holomorphic away from the locus \(Q(p, p) = 0\), while for odd \(n\) it contains a square root.  However, despite the apparent complication of this square root, these odd-$n$ integrals can be computed using the Gauss-Bonnet theorem for manifolds with corners. For instance, in the $n=3$ case, the edges of the triangle do not contribute since their geodesic curvature vanishes; correspondingly, only the vertices contribute. We will say more about this in section~\ref{sec:odd_n_integrals}.

%================================================================================================================
\vspace{-0pt}\subsection{Feynman Integrals as Hyperbolic Volumes}\label{subsec:feynman_integrals_as_volumes}\vspace{-0pt}
%================================================================================================================

Recall that in the projective model we have a projective space inhabited by points $v_i \in \mathbb{E}^{n-1,1}$ whose last components all equal unity, and a quadric defined by $Q(v_i, v_j) = 0$ whose points correspond to the boundary of hyperbolic space.  Consider an arbitrary point $I$ at infinity, namely a point satisfying \(Q(I, I) = 0\). All points \(v_i\) such that \(Q(I, v_i) = 0\) are also points at infinity, while points such that \(Q(I, v_i) \neq 0\) are points at finite distance. To each point not at infinity, we can associate another point 
\begin{align} \label{eq:massless_projection}
\hat{v}_i \equivR v_i + \lambda I\, , \qquad  \lambda = -\frac {Q(v_i, v_i)}{2 Q(v_i, I)}\,,
\end{align}
in which case we have that \(Q(I, \hat{v}_i) = Q(I, v_i)\).  Since \(\hat{v}_i\) is at infinity, it corresponds to an $n$-dimensional dual point. Thus, we can think of \(\hat{v}_i\) as a massless projection of \(v_i\), while \(\lambda\) parametrizes the protrusion of $v_i$ into the $n$-th dimension.

Given two such points \(v_i\) and \(v_j\), we define a set of four-dimensional distances and masses by
\eq{
  x_{i j}^2 \equivR -\frac {Q(\hat{v}_i, \hat{v}_j)}{Q(\hat{v}_i, I) Q(\hat{v}_j, I)}\,,\qquad  m_i^2 \equivR - \frac {Q(v_i, v_i)}{2 Q(v_i, I)^2}\,. \label{eq:mass_hyperbolic}
}
These quantities are invariant under the separate rescalings of \(\hat{v}_i\), \(\hat{v}_j\), and \(v_i\), while rescaling \(I\) should be thought of as a dilation transformation. It follows that
\begin{equation} \label{eq:Q_k_relation}
  \frac {-Q(v_i, v_j)}{\sqrt{-Q(v_i, v_i)} \sqrt{-Q(v_j, v_j)}} = \frac {x_{i j}^2 + m_i^2 + m_j^2}{2 m_i m_j} = \gram_{ij},
\end{equation}
where we have invoked the notation introduced in~\eqref{gram_matrix}. Plugging this relation into equation~\eqref{eq:projective-volume} and projectively rescaling $\alpha_i \mapsto \alpha_i/\sqrt{-Q(v_i, v_i)}$, we obtain 
\begin{align}
  \int\!\!d\text{vol}\left(\gram_{ij} \right)&= \frac 1 2 \int\!\!\!\proj{d^{n-1}\!\vec{\alpha}} \frac {\sqrt{|\det \gram|}}{\left(\sum_{i, j} \gram_{i j} \alpha_i \alpha_j \right)^{\frac n 2}}\,\\
  &= \frac{1}{2}\sqrt{\sigma(\gram)}\hat{I}_n(\gram) \, , \label{eq:feyn_int_is_simplex}
\end{align}
where $\hat{I}_n$ is the Feynman integral \eqref{I_hat} and $\sigma(\gram)$ was given in (\ref{sign_of_signature}). Thus, with the definitions \eqref{eq:mass_hyperbolic} we have an exact correspondence between volumes of $(n{-}1)$-simplices in hyperbolic space and one-loop $n$-particle Feynman integrals with arbitrary internal and external masses.

In order to invert relation~\eqref{eq:feyn_int_is_simplex} and express $\hat{I}_n$ (with a given set of internal masses and external momenta) as the volume of a simplex, recall that the hyperbolic distance $l_{ij}$ between two points $h_i$ and $h_j$ on the hyperboloid $\langle h_i, h_i \rangle = \langle h_j, h_j \rangle = -1$ was given in~\eqref{eq:hyperbolic_distance}, namely 
$-\left<h_i,h_j\right>=\cosh l_{ij}$.
In terms of the corresponding points in the projective model, $v_i$ and $v_j$, which form the same angle with respect to the origin of the ambient space (see Figure \ref{fig:hyperbolic_models}), this can be rewritten as ({\it c.f.} (\ref{eq:Q_k_relation}))
\begin{equation}
-\left<h_i,h_j\right>=\cosh l_{i j} = \frac {-Q(v_i, v_j)}{\sqrt{-Q(v_i, v_i)} \sqrt{-Q(v_j, v_j)}} = \gram_{ij}\,.\label{hyperbolic_lengths_and_gramian}
\end{equation}
Here we assume that all the off-diagonal entries of $\gram$ are greater than or equal to unity, so that this relation makes sense (we will discuss this point further in section~\ref{subsec:space-time_signatures}). To summarise, the matrix $\gram$ encodes the distances between all pairs of points forming the hyperbolic $(n-1)$-simplex we are after.  $\gram$ constitutes the (negative of the) Gram matrix\footnote{Named for the Danish mathematician J\o rgen Pedersen Gram, who met his demise in 1916 in the most Danish way imaginable: being struck by a bicycle in Copenhagen \cite{gram}.} of the corresponding points $h_i$ that define this simplex in the hyperboloid model, 
\begin{equation} \label{eq:hyperboloid_gram_matrix}
\langle h_i, h_j \rangle = \frac{\langle v_i, v_j\rangle}{\sqrt{-\langle v_i, v_i \rangle} \sqrt{-\langle v_j, v_j\rangle}} = \begin{pmatrix}
    -1 & -\cosh l_{1 2} & \hdots &  -\cosh l_{1 n} \\
    -\cosh l_{1 2} & -1 & \hdots &  -\cosh l_{2 n} \\
    \vdots & \vdots & \ddots & \vdots \\
    -\cosh l_{1 n} & -\cosh l_{2 n} & \hdots &  -1
  \end{pmatrix}.
\end{equation}
The lengths $l_{ij}$ uniquely specify a simplex in hyperbolic space up to isometries and therefore uniquely characterize the a simplicial volume. We can summarize this relation as stating that the Feynman integral $\hat{I}_n$ in (\ref{I_hat}) is given by 
\vspace{-4pt}
\eq{\begin{split}
\label{eqn:boxed_correspondance}
	\hat{I}_n = \sqrt{\sigma(\gram)}\,\,\text{vol}(l_{ij}) \,, \\[.2cm]
	\cosh l_{ij} = \gram_{ij} = \frac {x_{i j}^2 + m_i^2 + m_j^2}{2 m_i m_j} \,, 
\end{split}}
\noindent where $\sigma(\gram)$ was defined in (\ref{sign_of_signature}) and $\text{vol}(l_{ij})$ denotes the (unoriented) volume of a hyperbolic simplex in $n{-}1$ dimensions with edges of length $l_{ij}$, and these lengths satisfy \eqref{eqn:boxed_correspondance}. A similar set of variables \(r_{i j}\) were introduced in~\cite{Denner:1991qq}, which in our notation satisfy the relation
\begin{equation} \label{eq:r_vars}
\cosh l_{ij} = \frac{r_{i j} + r_{i j}^{-1}}{2} \, .
\end{equation}
It follows that \(r_{i j} = \exp l_{i j}\) if we choose the solution $r_{ij} > 1$.

%================================================================================================================
\vspace{-0pt}\subsection{{\it Exempli Gratia}: the Geometry of Hyperbolic Triangles}\label{subsec:geometry_of_hyperbolic_triangles}\vspace{-0pt}
%================================================================================================================

Unlike in Euclidean space, the volume of a hyperbolic simplex is uniquely determined by its angles. Thus, it is worth working out the relation between the lengths $l_{ij}$ and the dihedral angles $\phi^{(k)}_{ij}$ formed by the edges connecting vertices $h_i$ and $h_j$ with a third vertex $h_k$. We compute these angles in the hyperboloid model, where all vertices satisfy $\langle h_i, h_i \rangle = -1$.

\begin{figure}[t]\centering\fig{-10pt}{1}{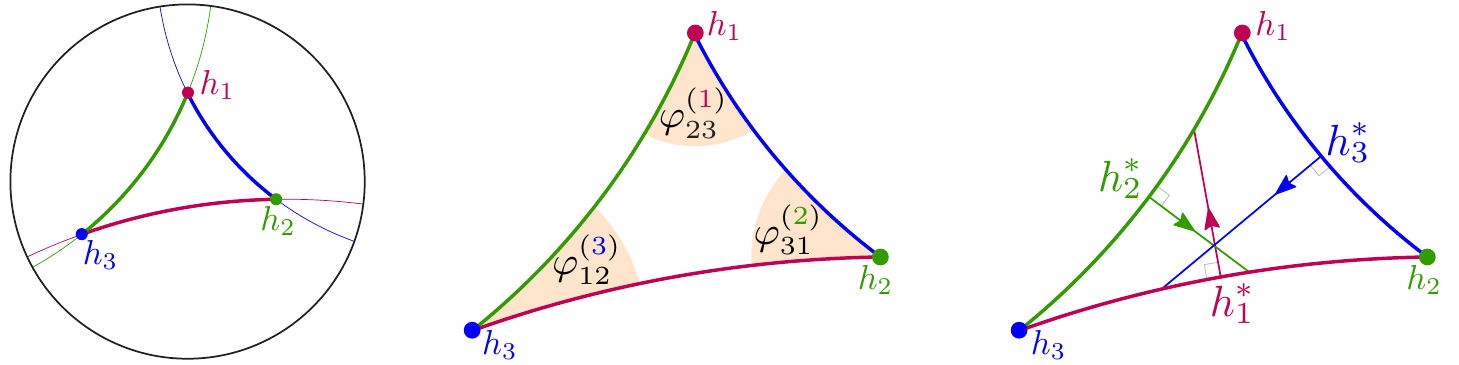}\caption{The vectors and angles defining a hyperbolic triangle formed by vertices $h_1$, $h_2$, and $h_3$.}\label{fig:hyperbolic_triangle_data}
\end{figure}

The vertices $h_1,h_2,h_3$ form a triangle with edge lengths given by $l_{12}$, $l_{13}$, and $l_{23}$, and we denote the angles opposite to these edges by $\smash{\phi^{(3)}_{12}}$, $\smash{\phi^{(2)}_{13}}$, and $\smash{\phi^{(1)}_{23}}$, as shown in \mbox{Figure \ref{fig:hyperbolic_triangle_data}}. We can also define this triangle by the three space-like vectors normal to its edges, $h^*_1$, $h^*_2$, and $h^*_3$, as shown there. The normalization of these vectors can be chosen so that they are dual to the original vectors $h_1$, $h_2$, and $h_3$, in the sense that 
\begin{equation} \label{eq:dual_vectors}
\langle h_i, h_j^* \rangle = \delta_{ij}.
\end{equation}
Note that this makes the vectors $h_j^*$ space-like. The dihedral angle between the two hyperplanes normal to $h^*_i$ and $h^*_j$ is the complement of the angle between these vectors, namely
\begin{equation}
  \phi^{(k)}_{ij} = \pi - \arccos  \left(\frac{\langle h^*_i, h^*_j \rangle}{\sqrt{\langle h^*_i, h^*_i \rangle} \sqrt{\smash{\langle h^*_j, h^*_j \rangle}\vphantom{\langle h^*_i, h^*_i \rangle}}} \right) \, ,  \label{eq:phi_complement}
\end{equation}
or equivalently
\begin{equation}
  \frac{\langle h^*_i, h^*_j \rangle}{\sqrt{\langle h^*_i, h^*_i \rangle} \sqrt{\smash{\langle h^*_j, h^*_j \rangle}\vphantom{\langle h^*_i, h^*_i \rangle}}} = -\cos \phi^{(k)}_{ij} \, . \label{}
\end{equation}
In these relations we have included square root factors that are equal to unity, as this will prove convenient below.

It follows from relation \eqref{eq:dual_vectors} that the Gram matrix of the dual vectors \(h_i^*\) is the inverse of the Gram matrix of \(h_i\) \eqref{eq:hyperboloid_gram_matrix}. Computing this, we find
\begin{align}
&\hspace{3cm} \langle h^*_i, h^*_j\rangle = 
     \begin{pmatrix}
    -1 & -\cosh l_{1 2} & -\cosh l_{1 3} \\
    -\cosh l_{1 2} & -1 & -\cosh l_{2 3} \\
    -\cosh l_{1 3} & -\cosh l_{2 3} & -1
  \end{pmatrix}^{-1}  \label{eq:dual_grammian}\\
  &\propto 
  \begin{pmatrix}
    \sinh^2 l_{2 3} & \cosh l_{1 2} - \cosh l_{1 3} \cosh l_{2 3} & \cosh l_{1 3} - \cosh l_{1 2} \cosh l_{2 3} \\
    \cosh l_{1 2} - \cosh l_{1 3} \cosh l_{2 3} & \sinh^2 l_{1 3} & \cosh l_{2 3} - \cosh l_{1 2} \cosh l_{1 3} \\
    \cosh l_{1 3} - \cosh l_{1 2} \cosh l_{2 3} & \cosh l_{2 3} - \cosh l_{1 2} \cosh l_{1 3} & \sinh^2 l_{1 2}
  \end{pmatrix}\, . \nonumber
\end{align}
Plugging the entries of this matrix into \eqref{eq:phi_complement}, we conclude that
\begin{align}
  \phi^{(3)}_{12} &= \arccos \left(\frac {-\langle h_1^*, h_2^*\rangle}{\sqrt{\langle h_1^*, h_1^*\rangle} \sqrt{\langle h_2^*, h_2^*\rangle}}\right) \nonumber \\
  &= \arccos \left(\frac {\cosh l_{1 3} \cosh l_{2 3} - \cosh l_{1 2}}{\sinh l_{1 3} \sinh l_{2 3}}\right). \label{eq:phi_3}
\end{align}
There exists a unique solution to this equation in the range \(\smash{0 < \phi^{(3)}_{12} < \pi}\). To see this, we assume without loss of generality that \(l_{2 3} \leq l_{1 3}\).  Then, the usual triangle inequality tells us that \(0 \leq l_{1 3} - l_{2 3} < l_{1 2} < l_{1 3} + l_{2 3}\).  Since the \(\cosh\) function is monotonically increasing on the positive real numbers, we have
\begin{equation}
  \cosh l_{1 3} \cosh l_{2 3} - \sinh l_{1 3} \sinh l_{2 3} < \cosh l_{1 2} < \cosh l_{1 3} \cosh l_{2 3} + \sinh l_{1 3} \sinh l_{2 3}.
\end{equation}
Rearranging these inequalities, we find
\begin{equation}
  -1 < \frac {\cosh l_{1 3} \cosh l_{2 3} - \cosh l_{1 2}}{\sinh l_{1 3} \sinh l_{2 3}} < 1.
\end{equation}
Since $\arccos$ is injective on this domain, this implies the value of \(\smash{0 < \phi^{(3)}_{1 2} < \pi}\) is unique. 

We can also invert relation~\eqref{eq:phi_3} (and the corresponding relations for $\phi^{(1)}_{23}$ and $\phi^{(2)}_{13}$) to compute the length $l_{12}$ in terms of the angles $\phi^{(k)}_{ij}$:
\begin{equation}
  \cosh l_{12} = \frac {\cos \phi^{(2)}_{13} \cos \phi^{(1)}_{23} + \cos \phi^{(3)}_{12}}{\sin \phi^{(2)}_{13} \sin \phi^{(1)}_{23}} \,. \label{eq:l_12}  
\end{equation}
Again, there exists a unique solution for $l_{12} >0$ whenever \(\phi^{(3)}_{1 2} + \phi^{(2)}_{1 3} + \phi^{(1)}_{2 3} < \pi\). Using the fact that \(\phi_{1 3}, \phi_{2 3} > 0\), we have \(0 < \phi_{1 2} < \pi - \phi_{1 3} - \phi_{2 3} < \pi\); since, moreover, the cosine decreases on the interval \([0, \pi]\),  
\begin{equation}
\cos \phi^{(3)}_{1 2} > \cos \left(\pi - \phi^{(2)}_{1 3} - \phi^{(1)}_{2 3} \right) = -\cos \phi^{(2)}_{1 3} \cos \phi^{(1)}_{2 3} + \sin \phi^{(2)}_{1 3} \sin \phi^{(1)}_{2 3} \, .
\end{equation}
Hence, \(\cosh l_{1 2} > 1\) and equation~\eqref{eq:l_12} has a single positive solution.

Rewriting relations~\eqref{eq:l_12} and~\eqref{eq:phi_3} for any triple of vertices $h_i$, $h_j$, and $h_k$, we have
\begin{gather}
  \cos \phi^{(k)}_{ij} = \frac {\cosh l_{ik} \cosh l_{jk} - \cosh l_{ij}}{\sinh l_{ik} \sinh l_{jk}} \, , \label{eq:phi_to_l} \\
  \cosh l_{ij} = \frac {\cos \phi^{(j)}_{ik} \cos \phi^{(i)}_{jk} + \cos \phi^{(k)}_{ij}}{\sin \phi^{(j)}_{ik} \sin \phi^{(i)}_{jk}} \, , \label{eq:l_to_phi} 
\end{gather}
where $\phi^{(k)}_{ij}$ is the angle formed between the edges emanating from $h_k$ to $h_i$ and $h_j$, and similarly for the other angles.  Note that when $\smash{\phi^{(k)}_{ij}}$ is a right angle, relation \eqref{eq:phi_to_l} reduces to the hyperbolic Pythagorean theorem
\begin{equation}
  \cosh l_{ik} \cosh l_{kj} = \cosh l_{ij} \, .
\end{equation}
Also, when the sides of the triangle are very small with respect to the radius of curvature of hyperbolic space (which we have taken to be 1), we obtain the usual Pythagorean theorem as an approximation.

%================================================================================================================
\vspace{-0pt}\subsection{Kinematic Domains and space-time Signatures}\label{subsec:space-time_signatures}\vspace{-0pt}
%================================================================================================================

Clearly, the interpretation of $\hat{I}_n$ as a volume in hyperbolic space will only be valid in certain kinematic regions; in particular, only for some values of $\gram_{ij}$ will the corresponding angles and lengths $\phi^{(k)}_{ij}$ and $l_{ij}$ be real numbers. Thus, we are led to ask: what are the constraints on \(\gram_{i j}\) such that a real hyperbolic simplex can be built from them?  

The answer to this question turns out to be related to the space-time signature in which we consider the integral $\hat{I}_n$. Consider a set of points $\{h_i\}$ with the Gram matrix \(\gram_{ij}=-\langle h_i, h_j\rangle\), where $\gram$ is given by some specific (but non-degenerate) choice of external momenta and masses. We can determine the signature \((n_+, n_-)\) of this kinematic point by finding a change of basis \(c_{i j}\) such that \(e_i = c_{i j}\, h_j\), with $\{c_{ij}\}$ real and where the \(e_i\) form the basis in which the scalar product is diagonal, \(\langle e_i, e_j\rangle = \pm \delta_{i j}\). The numbers $n_+$ and $n_-$ are then given by the number of positive and negative entries on the diagonal of $\langle e_i, e_j\rangle$, respectively. 

Consider, for instance, the signature of the Gram matrix encountered in the case of a hyperbolic triangle ($n=3$).  The characteristic polynomial of this matrix, which can be compactly expanded in powers of \(x + 1\), is
\begin{multline}
 - (x {+} 1)^3 + (\cosh^2 l_{1 2} + \cosh^2 l_{1 3} + \cosh^2 l_{2 3}) (x {+} 1)
  - 2 \cosh l_{1 2} \cosh l_{1 3} \cosh l_{2 3}.\label{eq:triangle_characteristic_poly}
\end{multline}
Computing the discriminant of this cubic equation in \(x + 1\) we find it to be
\begin{equation}
4 (\cosh^2 l_{1 2} + \cosh^2 l_{1 3} + \cosh^2 l_{2 3})^3 - 4 \times 27 \cosh^2 l_{1 2} \cosh^2 l_{1 3} \cosh^2 l_{2 3}, 
\end{equation}
which, due to the inequality between arithmetic and geometric means, must be positive. This implies that all the roots of this polynomial are real. 

Let us now assume that the space-time signature of our kinematic point is \((2,1)\), matching the scalar product~\eqref{eq:ambient_metric} of the ambient space $\mathbb{E}^{2,1}$. This implies that the product of the roots of~\eqref{eq:triangle_characteristic_poly} in the variable $x$ has to be negative:
\begin{equation} \label{eq:triangle_negative_prod_roots}
  - 2 \cosh l_{1 2} \cosh l_{1 3} \cosh l_{2 3} + \cosh^2 l_{1 2} + \cosh^2 l_{1 3} + \cosh^2 l_{2 3} - 1 < 0,
\end{equation}
where this inequality can be rewritten as
\begin{equation}
  (\cosh l_{1 2} - \cosh l_{2 3} \cosh l_{1 3})^2 < (\cosh^2 l_{1 3} - 1)(\cosh^2 l_{2 3} - 1) = \sinh^2 l_{1 3} \sinh^2 l_{2 3}.
\end{equation}
By comparison to equation \eqref{eq:phi_to_l}, we see that this condition implies $\smash{\cos^2 \left( \phi^{(3)}_{12} \right) < 1}$. Moreover, after extracting the square root and using the identity $\cosh a \cosh b + \sinh a \sinh b = \cosh (a + b)$, we also find the triangle inequality $l_{1 2} <  l_{1 3} + l_{2 3}$. The same reasoning can be applied to any orientation of the triangle, giving all three triangle inequalities and the same constraints on all three angles. We conclude that the correspondence \eqref{eqn:boxed_correspondance} is valid for $\hat{I}_3$ in all kinematic regions corresponding to $(2,1)$ signature. 

The converse of this statement also holds in general---that is, the Gram matrix of \(n\) vectors on the upper sheet of the hyperboloid in $\mathbb{E}^{n-1,1}$ must have signature \((n {-} 1, 1)\).  Any subset of \(k\) such vectors also generates a hyperbolic subspace, and hence their Gram matrix also has signature \((k {-} 1, 1)\). This is analogous to the situation in Euclidean space, where any \(n\) vectors of unit norm have signature \((n, 0)\), and any subset of \(k\) such vectors must similarly have signature \((k, 0)\).

For more general signatures there are more possibilities.  Consider \(n\) vectors with norm $-1$ in an embedding space of signature \((n{-}p,p)\). (We could equivalently take their norm to be \(1\), and exchange \(n{-}p \leftrightarrow p\).)  Given any subset of these vectors, we can compute the signature of their Gram matrix.  Which signatures are possible for the Gram matrices of all \(2^n\) possible subsets of the initial vectors?

There are two constraints these signatures must satisfy. First, the signature \((k{-}q,q)\) of any subset of $k$ vectors must satisfy \(k{-}q \leq n{-}p\) and \(q \leq p\). This immediately implies that the signature of all \(n\) vectors is the same as that of the embedding space. Second, whenever an additional vector is added to a subset of $k$ vectors with signature \((k{-}q, q)\), the resulting signature can only be \((k {-} q {+} 1, q )\) or \((k {-} q, q  {+} 1)\). To determine which it is, we project the new vector onto the orthogonal complement of the span of the original $k$ vectors. Whether this orthogonal projection has positive or negative norm tells us whether the new vector has increased the number of positive or negative eigenvalues of the Gram matrix.

More generally, in kinematic regions corresponding to signature $(n{-}p,p)$, the integral $\hat{I}_n$ can be interpreted as the volume of an $n$-simplex by taking $-\gram_{ij}$ to describe the Gram matrix of a set of $n$ vectors with norm $-1$ embedded in $\mathbb{E}^{n{-}p,p}$. Loosely, this corresponds to interpreting the entries of $-\gram_{ij}$ alternately as the cosine or the hyperbolic cosine of some angle, depending on whether the magnitude of the entry is greater than or less than unity. To reach such a region from regions of hyperbolic signature (where the correspondence~\eqref{eqn:boxed_correspondance} with all hyperbolic cosines holds) will in general require an intricate set of analytic continuations. However, the connection between the geometry of the $n$-simplex embedded in $\mathbb{E}^{n{-}p,p}$ and the external kinematics entering $\hat{I}_n$ should still be given by a projection of the simplicial vertices to the boundary of the hyperboloid on which these vertices lie, analogously to equations~\eqref{eq:massless_projection}--\eqref{eq:mass_hyperbolic}. For general $p$, the topology of this boundary (within the embedding space) will be given by a products of spheres \(\mathbb{S}^{n - p - 1} \times \mathbb{S}^{p - 1}\), where \(\mathbb{S}^{-1}\) should be interpreted as the empty set when $p$ equals $0$ or $n$.\footnote{As a consequence, no such boundary exists in the spherical signatures $(n,0)$ and $(0,n)$ for us to project onto. However, there is still a way to associate $\hat{I}_n$ with the volume of a simplex in these signatures~\cite{Davydychev:1997wa}. We leave an exploration of this point of view, which is valid in a general number of space-time dimensions, to a forthcoming companion paper~\cite{n_gons_d_dim}.}  Note that when \(p = 1\), we recover the hyperbolic case described in section~\ref{subsec:feynman_integrals_as_volumes}, where \(\mathbb{S}^{n - 1} \times \mathbb{Z}_2 = \mathbb{S}^{n - 1} \bigger{\cup} \mathbb{S}^{n - 1}\) corresponds to union of the \((n {-} 1)\)-dimensional spheres on the boundaries of the upper and lower branches of the hyperboloid.  

In other contexts, these regions with different space-time signatures have been seen to fit neatly together in real kinematics. For example, in four dimensions kinematic regions of signature \((3, 1)\) and \((2, 2)\) will be separated by a codimension-one boundary of signature \((2,1)\) along which all external momenta lie in a three-dimensional hypersurface. Along this boundary, quantities that are odd under space-time parity must vanish. This partitioning of kinematic space into regions of different signature can be nicely visualized when the number of kinematic variables is small, for instance in massless six-particle scattering in planar ${\cal N} = 4$ super-Yang-Mills theory~\cite{Alday:2009dv,Golden:2013xva,Dixon:2013eka,Dixon:2016apl}, which only depends on three kinematic invariants due to dual conformal symmetry~\cite{Drummond:2006rz, Bern:2006ew, Bern:2007ct, Alday:2007hr, Bern:2008ap, Drummond:2008vq}. This will also be the case for the bubble and triangle integrals we consider in the next section.  

We are unaware of the volumes of simplices being studied beyond the cases of Euclidean and hyperbolic (Lorentzian) signature, although functional representations of volumes that are valid in both of these signatures were considered in~\cite{MR1239859}. It would therefore be interesting to study volumes with ultra-hyperbolic signature.  In particular, it should be possible to extend the formula for the Euler characteristic that relates volumes in even dimensions to volumes in odd dimensions (which we discuss in section~\ref{sec:odd_n_integrals}) to these more general cases.

%================================================================================================================
%================================================================================================================
\newpage\vspace{-0pt}\section[All-Mass One-Loop Feynman Integrals in Low Dimensions]{All-Mass One-Loop Integrals in Low Dimensions}\label{sec:loop_integrals}\vspace{-0pt}
%================================================================================================================
As a warm-up, we first examine the correspondence between $n$-gons in $n$ dimensions and simplicial volumes for the cases of the bubble and the triangle. These integrals are simple enough that the results of direct Feynman integration can be straightforwardly compared to the corresponding hyperbolic volumes, providing a valuable cross-check on~\eqref{eqn:boxed_correspondance}. In this section, we also explore how the kinematic domains of these integrals are tiled by regions of different space-time signature, illustrating features of these integrals that we expect to hold for all $n$. 

%================================================================================================================
\vspace{-0pt}\subsection{The All-Mass Bubble Integral in Two Dimensions}\label{subsec:bubble}\vspace{-0pt}
%================================================================================================================
The simplest integral that has a hyperbolic volume interpretation is the one-loop massive bubble in two dimensions.  This integral depends on two internal masses, \(m_1\) and \(m_2\), and one external momentum.  From the Feynman integral representation~\eqref{conformal_int_ult} it can be easily evaluated to give
\begin{equation} \label{eq:bubble_integral}
  \hat{I}_2 = -i\,\sigma(\gram)\log r_{1 2}\equivR -i\sigma(\gram)\log\!\Big(\gram_{12}+\sqrt{\gram_{12}^2-1}\Big)\, ,
\end{equation}
where we have made use of the variables introduced in equation~\eqref{eq:r_vars}. Thus, $r_{12}$ corresponds to the larger of the two roots of the equation
\begin{equation} 
  \frac{1}{2}\Big(r_{1 2} + \frac 1 {r_{1 2}}\Big) = \frac {x_{1 2}^2 + m_1^2 + m_2^2}{2\,m_1\,m_2}=\gram_{12}\,;
\end{equation}
specifically, we require that $r_{12} > 1$ (in accordance with the argument of the logarithm in (\ref{eq:bubble_integral})). 

Let us now show that~\eqref{eq:bubble_integral} is precisely the volume of a simplex in \(\mathbb{H}^2\) whose geometry is determined by the kinematics of the two-point Feynman diagram depicted in Figure~\ref{fig:integral_and_dual}.  As per equations~\eqref{eq:massless_projection}-\eqref{eq:mass_hyperbolic}, the dual points $x_1$ and $x_2$ correspond to points on the boundary \(\partial \mathbb{H}^2\), while the internal masses $m_1$ and $m_2$ dictate how far from the boundary the two vertices of the corresponding hyperbolic simplex are located; in particular, a value of $m_i = 0$ implies that the $i^{\text{th}}$ simplicial vertex coincides with the dual point $x_i$ on \(\partial \mathbb{H}^2\).

The volume of a hyperbolic 1-simplex is just the length of the geodesic between its vertices, \(h_1\) and \(h_2\). From (\ref{eq:hyperboloid_gram_matrix}), this is just
\begin{equation} \label{eq:bubble_log}
  l_{1 2} = \arccosh (- \langle h_1, h_2\rangle) =
  \arccosh \gram_{12} =
  \log r_{1 2},
\end{equation}
matching the answer for $\hat{I}_2$ found through direct integration. Finally, we note that the massless limit of $\hat{I}_2$ is divergent when either of its propagators is massless. Geometrically, this corresponds to the corresponding simplicial vertex being sent to the boundary of $\mathbb{H}^2$, which causes the length of the geodesic to diverge.

%================================================================================================================
\vspace{3pt}\subsection{The All-Mass Triangle Integral in Three Dimensions}\label{subsec:triangle}\vspace{3pt}
%================================================================================================================
Let us now consider the triangle integral in three dimensions, which can be treated by the same methods. This integral was computed in~\cite{doi:10.1063/1.523697} using a judicious choice of cylindrical coordinates, and can be put in the form 
\begin{equation} \label{eq:feyn_triangle}
  \hat{I}_3(\gram) = 2 \arctan\!\left( \frac {\sqrt{\det \gram}}{1 + \gram_{12} + \gram_{13} + \gram_{23}} \right) \, .
\end{equation}
Note that $\arctan$ has unit transcendental weight and can be rewritten as a $\log$, but only at the expense of introducing imaginary arguments. 

\begin{figure}
  \centering
  \includegraphics[width=.4\textwidth]{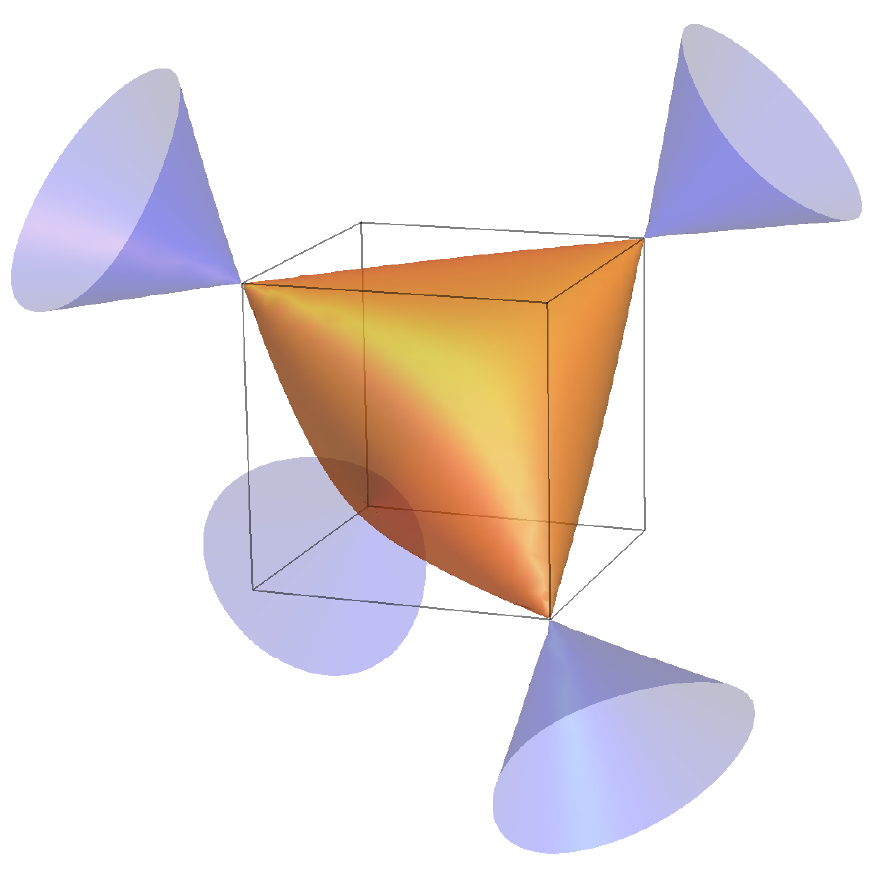}
  \caption{The boundary between regions of different space-time signature in triangle kinematics, as dictated by the inequality~\eqref{eq:triangle_regions}. The cube separating the inner and outer shaded regions marks the boundaries $\gram_{ij} = \pm 1$.}
  \label{fig:triangle_regions}
\end{figure}

We would again like to see that the same answer can be computed directly as a hyperbolic volume, which in this case is an area. But first, let us discuss the kinematic region in which this correspondence is expected to hold. Recasting inequality~\eqref{eq:triangle_negative_prod_roots} in terms of the kinematic variables $\gram_{ij}$, we have
\begin{equation} \label{eq:triangle_regions}
\det\gram=- 2 \gram_{1 2} \gram_{1 3} \gram_{2 3} + \gram_{1 2}^2 + \gram_{1 3}^2 + \gram_{2 3}^2 - 1 < 0,
\end{equation}
which must be satisfied whenever $\langle h_i, h_j \rangle = -\gram_{ij}$ has an odd number of negative eigenvalues. The surface where the left hand side of~\eqref{eq:triangle_regions} vanishes is plotted in Figure~\ref{fig:triangle_regions}. The inner (orange) region that this surface bounds must have signature $(0, 3)$, since at the origin $-\gram$ becomes proportional to the identity matrix. The unshaded region, which shares a codimension-one boundary with the inner region, has signature $(1,2)$. The remaining regions of kinematic space, shown in purple, have signature $(2,1)$, corresponding to the hyperbolic signature discussed in section~\ref{subsec:feynman_integrals_as_volumes}. The tiling of these regions exhibits a clear resemblance to the regions of different space-time signature encountered for six-particle scattering in planar $\mathcal{N} =4$ supersymmetric Yang-Mills theory (see for instance~\cite{Alday:2009dv,Dixon:2013eka}), although in that case there are no regions of spherical signature since the scattering particles are massless.

The area of a hyperbolic triangle is given by its angles as
\begin{equation} \label{eq:hyperbolic_triangle_area}
\pi - \phi^{(3)}_{12} - \phi^{(2)}_{13} - \phi^{(1)}_{23}.
\end{equation}
From equation~\eqref{eq:phi_to_l} and the identification of $\cosh l_{ij}$ with $\gram_{ij}$ we have
\begin{equation} \label{eq:phi_to_G}
  \cos \phi^{(k)}_{i j} = \frac {\gram_{i k} \gram_{j k} - \gram_{i j}}{\sqrt{\gram_{i k}^2 - 1} \sqrt{\smash{\gram_{j k}^2 - 1} \vphantom{\gram_{i k}^2 - 1}}} \, .
\end{equation}
Using the identity \(\smash{\arccos a = \arctan \left( \frac {\sqrt{1 - a^2}} a\right)} \) and the fact that $\gram_{ij}>1$ in this region, we can express $\smash{\phi^{(k)}_{ij}}$ as 
\begin{equation}\label{phi_in_arctan}
  \phi^{(k)}_{ij} = \arctan \left( \frac {\sqrt{\det \gram}}{\gram_{i k} \gram_{j k} - \gram_{i j}} \right).
\end{equation}
Next we substitute (\ref{phi_in_arctan}) into (\ref{eq:hyperbolic_triangle_area}) and demonstrate the the latter reproduces the triangle result of (\ref{eq:feyn_triangle}). Knowing that we need to cancel off the factor of $\pi$ in~\eqref{eq:hyperbolic_triangle_area}, we invert the arctangent's arguments in two of the angles using \(\smash{\arctan a = \frac {\pi} 2} - \arctan \frac 1 a \). After combining everything into a single term using $\smash{\arctan a \pm \arctan b = \arctan \big(\frac{a\pm b}{1 \mp a b} \big)}$, the identity \(\smash{\arctan \big( \frac {2 a}{1 - a^2} \big) = 2 \arctan a}\) allows us to reproduce~\eqref{eq:feyn_triangle} as desired.

In fact, the same expression is also valid in the spherical region corresponding to $(0,3)$ space-time signature. As can be seen in Figure~\ref{fig:triangle_regions}, this region intersects the hyperbolic region considered above at the point $\gram_{12}=\gram_{13}=\gram_{23}=1$; thus, we can analytically continue into spherical signature along the line $\gram_{12}=\gram_{13}=\gram_{23}=z$. Rewriting~\eqref{eq:feyn_triangle} as a logarithm and restricting to this line, we have
\begin{equation}
\hat{I}_3\left(\gram \big|_{\gram_{12}=\gram_{13}=\gram_{23}=z} \right) = i \log \left( \frac{(1+3z)- i \sqrt{(z-1)^2(1+2z)}}{(1+3z)+i \sqrt{(z-1)^2(1+2z)}} \right) \, ,
\end{equation}
which is valid both in the hyperbolic region $z>1$ and the Euclidean region $z<1$. To see this, notice that no imaginary part will be generated when we analytically continue into the spherical region $z < 1$ no matter which way we continue \mbox{$(z-1) \to e^{\pm i \pi}|(z-1)|$}. The net effect, with either choice, is to flip the signs in front of the square roots, inverting the argument of the logarithm. When considered beyond this particular line through kinematic space, the only alteration can arise as a phase due to $\sqrt{\sigma(\gram)}$. Thus, we may conclude that 
\begin{equation} \label{eq:spherical_triangle}
2 \arctan\!\left(\frac {\sqrt{\det \gram}}{1 + \gram_{1 2} + \gram_{1 3} + \gram_{2 3}}\right) \equivL V_3(\gram) \sqrt{\sigma(\gram)}  \, ,
\end{equation}
holds in every signature. Notice that we  have adopted the notation (both here and below) that $V_n(\gram)$ denotes the volume of an $(n-1)$-dimensional simplex in spherical signature that has edges of length $\gram_{ij} = \cos l_{ij}$.

Note that if we run the trigonometric argument below~\eqref{eq:hyperbolic_triangle_area} in reverse while using $\gram_{ij} = \cos l_{ij}$ to define a set of edge lengths,~\eqref{eq:spherical_triangle} can be understood as giving the area of a spherical triangle with angles $\phi^{(k)}_{ij}$:
\begin{equation}
-\pi +\phi^{(3)}_{12} + \phi^{(2)}_{13} + \phi^{(1)}_{23} \quad \text{(mod $4\pi$)}\, . \label{vol_spherical_triangle}
\end{equation}
This differs from the area for a hyperbolic triangle~\eqref{eq:hyperbolic_triangle_area} only by an overall sign. This area is interpreted modulo $4\pi$ since the area of a spherical triangle cannot be larger than the area of the sphere in which it's embedded.

%================================================================================================================
\vspace{-0pt}\section{The All-Mass Box Integral in Four Dimensions}\label{subsec:box}\vspace{-0pt}
%================================================================================================================

Let us now consider the all-mass box integral in four dimensions. In kinematic regions with space-time signature (3,1), this integral will be given by the volume of a hyperbolic tetrahedron formed by four vertices $h_i$ in $\mathbb{H}^3$. This kinematic region is picked out by five conditions in addition to our usual requirement that $\gram_{ij} \ge 1$. Four inequalities come from the requirement that the codimension-one faces of the tetrahedron form hyperbolic triangles---that is, the requirement that~\eqref{eq:triangle_regions} be satisfied for any choice of three of the four vertices $h_i$. As per the discussion in section~\ref{subsec:space-time_signatures}, once these constraints are satisfied the Gram matrix of the full tetrahedron can only have space-time signature $(3,1)$ or $(2,2)$. The last constraint is thus supplied by the requirement that the product of all four eigenvalues of $\gram$ be negative, namely $\det\gram < 0$. Note that this last requirement ensures that the normalization of~\eqref{I_hat}, $\sqrt{\det\gram}$, is purely imaginary. 

%================================================================================================================
\vspace{-0pt}\subsection{The Murakami-Yano Formula}\label{subsec:box_in_angles}\vspace{-0pt}
%================================================================================================================

A concise formula for the volume of a hyperbolic tetrahedron was given by Murakami and Yano in~\cite{MR2154824}. To present this formula, we define a set of dual vectors $h_i^*$ by the orthogonality condition
\begin{equation} \label{eq:box_dual_vectors}
\langle h_i, h_j^* \rangle = \delta_{ij}\,,
\end{equation}
just as we did for the hyperbolic triangle in section~\ref{subsec:geometry_of_hyperbolic_triangles}. Importantly, these space-like vectors encode the full geometry of the tetrahedron; in particular, its codimension-one faces (the hyperbolic triangles formed out of any three of the tetrahedron's vertices) are each orthogonal to one of these dual vectors (namely, the vector dual to the fourth tetrahedron vertex). The dihedral angles between these faces are thus encoded in the angles between the dual vectors. 

To compute these angles for a tetrahedron described by the Gram matrix $-\gram$, we rescale the rows and columns of $\gram^{-1}$ (in a manner that keeps it symmetric) so that the resulting matrix has diagonal entries equal to $-1$. This defines for us a matrix $\gram^*$ with entries 
\eq{\gram^*_{ij}\equivR\frac{\gram^{-1}_{ij}}{\sqrt{\gram^{-1}_{ii}}\sqrt{\smash{\gram^{-1}_{jj}}\vphantom{\gram^{-1}_{ii}}}}\equivL-\cos\theta_{ij}\,,\label{gstar_defined}}
where our notation is such that `$\gram^{-1}_{ij}$' denotes a component of the matrix $\gram^{-1}$. The angle $\theta_{ij}$ defined in the last step gives the angle between the dual vectors $h^*_i$ and $h^*_j$. 

In hyperbolic signature, the angles $\theta_{ij}$ are guaranteed to be real; as such, it is natural to define a set of phases
\eq{\begin{array}{r@{}lr@{}lr@{}l}a &\equivR e^{i \theta_{12}}\,, \qquad&b &\equivR e^{i \theta_{13}}\,, \qquad&c& \equivR e^{i \theta_{23}}\,,\\
d&\equivR e^{i \theta_{34}}\,,&e&\equivR e^{i \theta_{24}}\,,&f&\equivR e^{i \theta_{14}}\,.
\end{array}\label{abcdef_defined}}
Finally, we define a weight-two function
\eq{\begin{split}U(z) \equivR&\hspace{-2pt}\phantom{-}\li_2(z) + \li_2(a b d e z) + \li_2(a c d f z) + \li_2(b c e f z)\\&\hspace{-2pt}\!\!-\li_2(-a b c z) - \li_2(-a e f z) - \li_2(-b d f z) - \li_2(-c d e z)\,\label{u_function_defined}
\end{split}}
and a pair of roots
\eq{z_\pm \equivR -2 \frac {\sin \theta_{12} \sin \theta_{34} + \sin \theta_{13} \sin \theta_{24} + \sin \theta_{23} \sin \theta_{14} \pm \sqrt{\det \gram^*}}{a d + b e + c f + a b f + a c e + b c d + d e f + a b c d e f}\,. \label{zpm_for_murakami_yano_defined}}
The volume of the designated tetrahedron is then given by
\eq{\mathrm{vol}\big(\gram\big)=\frac{1}{4}\Im\Big[U(z_+)-U(z_-)\Big]\,,}
where $\Im$ denotes the imaginary part. This renders the (kinematic part of the) all-mass box in four dimensions to be
\eq{\hat{I}_4\big(\gram\big)=\sqrt{\sigma(\gram)}\,\mathrm{vol}\big(\gram\big)=\sqrt{\sigma(\gram)}\frac{1}{4}\Im\Big[U(z_+)-U(z_-)\Big]\,\label{murakami_yano_formula_for_box}}
due to the normalization for $I_4$ chosen in (\ref{conformal_int_ult}).

The Murakami-Yano expression for the all-mass box~\eqref{murakami_yano_formula_for_box} agrees with those already found in the physics literature (see for example~\cite{Denner:1991qq,Davydychev:2017bbl}), but has several remarkable features that make it distinct. In addition to the manifest simplicity of (\ref{murakami_yano_formula_for_box}), it exhibits full permutation invariance among all four hyperbolic vertices, and correspondingly in the external particles' dual-momentum variables' indices. This symmetry amounts to an invariance of $\hat{I}_4(\gram)$ under permutations of the rows and columns $\gram_{i\,j}\mapsto\gram_{\sigma(i)\,\sigma(j)}$ for any $\sigma\in\mathfrak{S}_4$. To see this, it is sufficient to notice that $z_+$ and $z_-$ are separately invariant, and the arguments of the dilogarithms in (\ref{u_function_defined}) form a three-orbit $\{a b d e\, z, a c d f\, z, b c e f\, z\}$ and four-orbit $\{-a b c\, z, -a e f\, z, -b d f\, z, -c d e\, z\}$. (Given the invariance of $z_\pm$, these orbits are easy to identify from the index structure defining the phases (\ref{abcdef_defined}).)

%================================================================================================================
\vspace{-0pt}\subsection{The All-Mass Box in Euclidean Signature}\label{subsec:box_euclidean_signature}\vspace{-0pt}
%================================================================================================================

It turns out that Murakami has also given a compact formula for the volume of a tetrahedron in spherical signature~\cite{MR2917101}. This formula makes use of the angular variables introduced in~\eqref{abcdef_defined}, but requires the (positive-root) solution $\zeta_+$ of the quadratic $q_2\zeta^2+q_1\zeta+q_0=0$, where
\eq{\begin{split}
  q_0 &\equivR a d + b e + c f + a b f + a c e + b c d + d e f + a b c d e f, \\
  q_1 &\equivR -(a - 1/a) (d - 1/d) - (b - 1/b) (e - 1/e) - (c - 1/c) (f - 1/f), \\
  q_2 &\equivR (a d)^{-1} + (b e)^{-1} + (c f)^{-1} + (a b f)^{-1} + (a c e)^{-1} \\
  &\hspace{137.85pt}+ (b c d)^{-1} + (d e f)^{-1} + (a b c d e f)^{-1}.
\end{split}}
We also require the function
\eq{\begin{split}
  L(\zeta) \equivR& \frac 1 2 \Big[\phantom{-}\li_2\!(\zeta) +
            \li_2\!\Big(\frac{\zeta}{a b d e}\Big) +
            \li_2\!\Big(\frac{\zeta}{a c d f} \Big) +
            \li_2\!\Big(\frac{\zeta}{b c e f} \Big) \\
           &\hspace{14pt}\!-\! \li_2\!\Big(-\frac{\zeta}{a b c}\Big) -
            \li_2\!\Big(-\frac{\zeta}{a e f}\Big) -
            \li_2\!\Big(-\frac{\zeta}{b d f}\Big) -
            \li_2\!\Big(-\frac{\zeta}{c d e}\Big) \\
            &\hspace{14pt}\!+\!\log(a) \log(d) + \log(b) \log(e) + \log(c) \log(f)
\Big].
\end{split}}
In terms of $L(\zeta_+)$, the volume of the spherical tetrahedron is given by
\eq{%
  V_4(\gram)=  -\Re\left(L(\zeta_+)\right) + \pi \Bigl(\arg (-q_2) + \frac 1 2 \sum_{i < j} \theta_{i j}\Bigr) - \frac {3 \pi^2}{2} \quad \text{(mod $2\pi^2$)} \,.\label{eq:spherical_tetrahedron_vol}}
Like with the formula for the volume of a spherical triangle,~\eqref{vol_spherical_triangle}, this formula is only valid modulo \(2 \pi^2\) because the volume of a tetrahedron embedded in a four-sphere cannot be larger than the volume of the sphere itself. It can be checked that $\hat{I}_4\big(\gram\big) = V_4(\gram)$ in this region, as expected. This formula also makes manifest the permutation invariance of this integral, in the same way as was observed in~\eqref{murakami_yano_formula_for_box}.

%================================================================================================================
\vspace{-0pt}\subsection{Recasting Murakami-Yano from Angles to `Lengths'}\label{subsec:box_in_lengths}\vspace{-0pt}
%================================================================================================================

While equations~\eqref{murakami_yano_formula_for_box} and~\eqref{eq:spherical_tetrahedron_vol} exhibit remarkable simplicity, one reasonable complaint about them is the sheer definitional distance between our kinematic variables (the Mandelstams $x_{ij}^2$ and masses $m_i^2$) and the angular variables appearing in the dilogarithms, logarithms, and roots. The algebraic complexities involved in these definitions pose no problem for numeric evaluation, but obfuscate the physically-relevant analytic structure of the all-mass box. This can be remedied by fully unpacking the definitions (\ref{abcdef_defined}) and (\ref{zpm_for_murakami_yano_defined}), and simplifying what emerges.

As we have already seen (for instance in equation~\eqref{eq:bubble_integral}), it can be a good idea to use hyperbolic `length-like' variables to describe the kinematic variables in $\gram$. Specifically, we might want to recast the Mandelstam invariants $x_{ij}^2$ and internal masses $m_i^2$ in terms of the $r_{ij}$ variables defined in~\eqref{eq:r_vars}. From the definition of $\gram^*$ in (\ref{gstar_defined}), the angular variables in (\ref{abcdef_defined}) can be expressed as
\eq{\begin{array}{ll}
\displaystyle a=\frac{1}{\sqrt{\gram^{-1}_{11}\gram^{-1}_{22}}}\Big(\gram^{-1}_{12}+\frac{\sinh l_{34}}{\sqrt{\det\gram}}\Big)\,,\qquad &\displaystyle b=\frac{1}{\sqrt{\gram^{-1}_{11}\gram^{-1}_{33}}}\Big(\gram^{-1}_{13}+\frac{\sinh l_{24}}{\sqrt{\det\gram}}\Big)\,,\\
\displaystyle c=\frac{1}{\sqrt{\gram^{-1}_{22}\gram^{-1}_{33}}}\Big(\gram^{-1}_{23}+\frac{\sinh l_{14}}{\sqrt{\det\gram}}\Big)\,,\qquad &\displaystyle d=\frac{1}{\sqrt{\gram^{-1}_{33}\gram^{-1}_{44}}}\Big(\gram^{-1}_{34}+\frac{\sinh l_{12}}{\sqrt{\det\gram}}\Big)\,,\\
\displaystyle e=\frac{1}{\sqrt{\gram^{-1}_{22}\gram^{-1}_{44}}}\Big(\gram^{-1}_{24}+\frac{\sinh l_{13}}{\sqrt{\det\gram}}\Big)\,,\qquad &\displaystyle f=\frac{1}{\sqrt{\gram^{-1}_{11}\gram^{-1}_{44}}}\Big(\gram^{-1}_{14}+\frac{\sinh l_{23}}{\sqrt{\det\gram}}\Big)\,,\\
\end{array}\label{eq:box_angles_as_lengths}}
where $\sinh l_{ij}=\frac{1}{2}(r_{ij}-1/r_{ij})=\sqrt{\gram_{ij}^2-1}$ and $\gram^{-1}_{ij}\equivR\big(\gram^{-1}\big)_{ij}$ are elements of the inverse of $\gram$ as before. In terms of these variables, one might expect that the arguments of the polylogarithms appearing in (\ref{u_function_defined}) would involve lengthy algebraic expressions (arising from the inverse matrix elements) as well as many algebraic roots. It turns out that this is not the case. In fact, when $\gram$ is expressed in terms of the $r_{ij}$, the \emph{only} algebraic root appearing in any of the arguments of the polylogarithms of $U(z)$ will be $\sqrt{\det\gram}$. 

As discussed above, the function $U(z)$ can be generated as a sum over three orbits which permute the rows and columns of $\gram_{ij}$. Thus, it suffices for us to give three of these expressions, and generate the rest via relabelings. We therefore consider the following three arguments of dilogarithms in $U(z_-)$ as defined by (\ref{u_function_defined}),
\eq{g_0(r_{ij})\equivR z_-\,,\;\;\;\;\;\; g_1(r_{ij})\equivR a b d e\, z_-\,,\;\;\;\;\;\; g_2(r_{ij})\equivR -a b c\,z_-\,,\label{gis_defined}}
where we note again that all of the square roots in~\eqref{eq:box_angles_as_lengths} other than $\sqrt{\det\gram}$ appear in pairs and drop out. Thus, these functions involve the single algebraic root
\eq{\delta\equivR 4(r_{12}r_{13}r_{14}r_{23}r_{24}r_{34})\sqrt{\det\gram}\,,}
where we have introduced this notation because $\delta^2$ will be a \emph{polynomial} in the $r_{ij}$ variables with integer coefficients.

In terms of $\delta$, the arguments of the polylogarithms $g_0(r_{ij})$, $g_1(r_{ij})$, and $g_2(r_{ij})$ can be compactly expressed as
\eq{g_0\equivR1+\frac{\delta}{\rho\,y_0}\big(\delta+x_0\big)\,,\;\;\;\;\;\;g_1\equivR1+\frac{\delta}{\rho\,y_1}\big(\delta+x_1\big)\,,\;\;\;\;\;\;g_2\equivR1+\frac{\delta}{\rho\,y_2}\big(\delta+x_2\big)\,,}
where $\rho$, $y_i$, and $x_i$ are given by
\eq{\begin{split}~\hspace{-100pt}y_0&\equivR\rs{123}\rs{124}\rs{134}\rs{234}\,,\;\;\;\;\;\;\;\;y_1\equivR r_{12}r_{24}r_{43}r_{31}\,\ru{1}{23}\ru{2}{14}\ru{3}{41}\ru{4}{32}\,,\;\;\;\;\;\;\;\;y_2\equivR r_{123}\rs{123}\,\ru{4}{12}\ru{4}{23}\ru{4}{31}\,,\hspace{-40pt}\\
~\hspace{-100pt}x_0&\equivR \rho+(r_{12}r_{13}r_{14}r_{23}r_{24}r_{34})\Big(\frac{r_{12}}{r_{34}}\!+\!\frac{r_{13}}{r_{24}}\!+\!\frac{r_{14}}{r_{23}}\!+\!\frac{r_{23}}{r_{14}}\!+\!\frac{r_{24}}{r_{13}}\!+\!\frac{r_{34}}{r_{12}}\!-\!r_{12}r_{34}\!-\!r_{14}r_{23}\!-\!r_{13}r_{24}\Big)\hspace{-40pt}\\
&\hspace{21pt}-\big(r_{12}r_{23}r_{34}r_{41}+r_{13}r_{34}r_{42}r_{21}+r_{14}r_{42}r_{23}r_{31}\big)\,,\\
\hspace{-100pt}x_1&\equivR x_0+2\big(1-r_{12}r_{13}r_{24}r_{34}\big)\big(\rs{123}-\ru{1}{24}r_{23}r_{34}-\ru{2}{14}r_{13}r_{34}-\ru{3}{12}r_{14}r_{24}\big)\,,\hspace{-40pt}\\
~\hspace{-100pt}x_2&\equivR x_0+2\,\rs{123}\Big(\!1\!-\!r_{\!124}\!-\!r_{\!134}\!-\!r_{\!234}+\ru{4}{12}\ru{4}{13}\ru{4}{23}+r_{\!14}r_{\!24}r_{\!34}\big(r_{\!12}r_{\!34}+r_{\!23}r_{41}+r_{13}r_{24}\big)\hspace{-40pt}\\
&\hspace{75.75pt}-\big(r_{14}r_{24}r_{34}\big)^2\Big)\,,\\
~\hspace{-100pt}\rho&\equivR2\Big(\!1\!-\!\big(r_{\!123}+r_{\!124}+r_{\!134}+r_{\!234}\big)+\big(r_{12}r_{23}r_{34}r_{41}+r_{13}r_{34}r_{42}r_{21}+r_{14}r_{42}r_{23}r_{31}\big)\!\Big)\,,\hspace{-40pt}
\end{split}}
where we have made use of the short-hand
\eq{r_{i j k}\equivR r_{i j} r_{j k} r_{k i}\,,\;\;\;\;\;\;\;\;\;\rs{i j k}\equivR r_{i j k}-1\,,\;\;\;\;\;\;\;\;\; \ru{i}{j k}\equivR r_{i j} r_{i k}-r_{j k}\,.} 
Notice that $\rho$, $y_0$, and $x_0$ are each invariant under arbitrary permutations of the rows and columns of $\gram_{ij}$, making the invariance of $g_0(r_{ij})$ under these transformations manifest. 

To make clear how the full set of arguments in~\eqref{u_function_defined} is generated from the three in~\eqref{gis_defined}, we denote the images of $g_k(r_{ij})$ under permutations $\sigma\!\in\!\mathfrak{S}_4$ by
\eq{g_{k}^{\sigma}\equivR g_{k}\left(r_{ij} \big|_{i,j\to\sigma(i),\sigma(j)}\right)\quad\text{and write}\quad g_{k}^{\sigma}\equivR g_{k}^{\sigma(1)\cdots\sigma(4)}\,.}
The function $U(z_-)$ is then given by:
\eq{\begin{split}
-U(z_-)=&\phantom{-}\li_2\!\big(g_0^{1234}\big)+\li_2\!\big(g_1^{1234}\big)+\li_2\!\big(g_1^{1342}\big)+\li_2\!\big(g_1^{1423}\big)\\
&\hspace{0pt}\!-\!\li_2\!\big(g_2^{1234}\big)-\li_2\!\big(g_2^{2341}\big)-\li_2\!\big(g_2^{3412}\big)-\li_2\!\big(g_2^{4123}\big)\,.
\end{split}\label{uzminus_formula}}
What about $U(z_+)$? In $(3,1)$ signature, it turns out that $z_-\leftrightarrow z_+$ is generated by $r_{ij}\leftrightarrow1/r_{ij}$ together with complex conjugation; in particular,
\begin{align} \label{eq:z_plus_from_minus_hypebolic}
z_+=g_0^*\!\big(1/r_{ij}\big)\,, \qquad a b d e\, z_+=g_1^*\!\big(1/r_{ij}\big)\,, \qquad -a b c\, z_+=g_2^*\!\big(1/r_{ij}\big)\,, 
\end{align}
where `$*$' denotes complex conjugation. In this signature, complex conjugation just amounts to changing the sign of $\sqrt{\det \gram}$ (when the $r_{ij}$'s are all real). 

The clever reader may notice that (\ref{murakami_yano_formula_for_box}) involves only the imaginary parts of $U(z_{\pm})$ and be tempted to simply add to (\ref{uzminus_formula}) the same expression with $r_{ij}\leftrightarrow1/r_{ij}$ exchanged. This will indeed yield the correct imaginary part to reproduce $\hat{I}_4$ in this signature. However, it turns out to be better to keep the conjugation inside the arguments (as we will thereby derive a formula with much greater validity). Specifically, let us define
\eq{\bar{g}_k(r_{ij})\equivR\left.\Big(1+\frac{\delta}{\rho\,y_k}(\delta-x_k)\Big)\right|_{r_{ij}\mapsto1/r_{ij}}\,,}
and consider the branch choice of $\delta$ to be the same for all $g_i$ and $\bar{g}_i$. This reproduces~\eqref{eq:z_plus_from_minus_hypebolic} in $(3,1)$ signature, but it turns out to hold more generally. Given this definition, (\ref{murakami_yano_formula_for_box}) can be put in the form 
\begin{align}
\hspace{-10pt}\hat{I}_4(r_{ij})&=\sqrt{\sigma(\gram)}\frac{1}{4}\Im\Big[\li_2\!\big(g_0^{1234}\big)+\li_2\!\big(g_1^{1234}\big)+\li_2\!\big(g_1^{1342}\big)+\li_2\!\big(g_1^{1423}\big)\label{all_mass_box_r_formula}\\
&\hspace{25.35pt}-\li_2\!\big(g_2^{1234}\big)-\li_2\!\big(g_2^{2341}\big)-\li_2\!\big(g_2^{3412}\big)-\li_2\!\big(g_2^{4123}\big)-\Big(g_{i}^{\sigma}\leftrightarrow\bar{g}_i^{\sigma}\Big)\Big]\,,\nonumber
\end{align}
Remarkably enough, it turns out that (\ref{all_mass_box_r_formula}) holds in all space-time signatures(!). We have checked this explicitly at many randomly chosen kinematic points with signatures $(4,0)$, $(3,1)$, and $(2,2)$. Before moving on, we should mention that a different and intriguing version of the Murakami-Yano formula expressed in terms of lengths should follow from the work of \cite{2019arXiv190801141R}; it would be worthwhile to see how these compare.
% 

%================================================================================================================
\vspace{-0pt}\subsection{A Dihedrally-Invariant Kinematic Limit}\label{subsec:box_diehdral_limit}\vspace{-0pt}
%================================================================================================================

The all-mass integral is symmetric under arbitrary permutations of the dual coordinates \((x_i, m_i)\). However, there are a number of contexts in which one just wants dihedral invariance in physics---for example, in the context of dual conformal (and ultimately Yangian) symmetry in planar integrals in maximally supersymmetric Yang-Mills theory \cite{Drummond:2006rz,Drummond:2007au,Drummond:2007aua,Drummond:2008vq,Alday:2009yn,Bourjaily:2013mma}. 

One such (dihedrally invariant) limit was introduced in the so-called `Higgs' regularization scheme described in~\cite{Alday:2009zm,Henn:2010bk} (see also \cite{Hodges:2010kq}). Here, one considers general masses for propagators around the perimeter of the graph in a planar ordering. Taking the points $\{x_1,x_2,x_3,x_4\}$ to be cyclically ordered, one then imposes a `five-dimensional on-shell' condition of the form:
\eq{x_{i,i+1}^2+(m_i-m_{i+1})^2=0\,.\label{dihedral_condition}}
Considering the definition of $\gram_{ij}$, it is easy to see that $\gram_{i,i+1}\mapsto1$ in this limit:
\eq{\gram\underset{\text{(\ref{dihedral_condition})}}{\longmapsto}\left(\begin{array}{@{}c@{}c@{}c@{}c@{}}1&1&\gram_{13}&1\\1&1&1&\gram_{24}\\\gram_{13}&1&1&1\\1&\gram_{24}&1&1\end{array}\right)\,.}
In terms of the variables $u$ and $v$ introduced in~\cite{Henn:2010bk}, namely
\eq{4u\equivR\frac{m_1m_3}{x_{13}^2+(m_1-m_3)^2}\qquad\text{and}\qquad 4v\equivR\frac{m_2m_4}{x_{24}^2+(m_2-m_4)^2}\,,\label{defn_of_uv_higgs_reg}}
the Gram matrix above takes the form
\eq{\gram\underset{\substack{\text{(\ref{dihedral_condition})}\\\text{(\ref{defn_of_uv_higgs_reg})}}}{\longmapsto}\left(\begin{array}{@{}c@{}c@{}c@{}c@{}}1&1&1+\frac{2}{u}&1\\1&1&1&1+\frac{2}{v}\\1+\frac{2}{u}&1&1&1\\1&1+\frac{2}{v}&1&1\end{array}\right)\,.}
Notice that in terms of these variables, $\sqrt{\det\gram}=\frac{4}{u v}\sqrt{1+u+v}$, and we can choose the corresponding variables $r_{ij}$ to be
\eq{r_{13}\equivR1+2\big(1+\sqrt{1+u}\big)/u\qquad\text{and}\qquad r_{24}\equivR1+2\big(1+\sqrt{1+v}\big)/v\,}
while all other $r_{ij} = 1$.

In this limit, the formula for $\hat{I}_4$ simplifies considerably. In particular, symmetry considerations allow us to identify 
\eq{g_0^{1234}=g_1^{1423}\,,\hspace{20pt}
g_1^{1234}=g_1^{1342}\,,\hspace{20pt}
g_2^{1234}=g_2^{3412}\,,\hspace{20pt}
g_2^{2341}=g_2^{4123}\,.
}
This collapses the 16-term formula for $\hat{I}_4(r_{ij})$ in (\ref{all_mass_box_r_formula}) to 
\begin{align}
\hspace{-10pt}\left.\hat{I}_4(r_{13},r_{24})\right|_{r_{i,i+1}=1}\hspace{-12pt}&=\sqrt{\sigma(\gram)}\frac{1}{2}\Im\Big[\li_2\!\big(g_0^{1234}\big)+\li_2\!\big(g_1^{1234}\big)-\li_2\!\big(g_2^{1234}\big)-\li_2\!\big(g_2^{2341}\big)
\label{dihedral_all_mass_box_r_formula}\\
&\hspace{60.95pt}-\li_2\!\big(\bar{g}_0^{1234}\big)-\li_2\!\big(\bar{g}_1^{1234}\big)+\li_2\!\big(\bar{g}_2^{1234}\big)+\li_2\!\big(\bar{g}_2^{2341}\big)\Big]\,,\nonumber
\end{align}
which is considerably more compact.

It is interesting to note that there is essentially no difference between the limit we have just considered---in which there are four unequal internal masses while the external momenta are constrained by (\ref{dihedral_condition})---and the more familiar kinematic limit in which all internal masses are equal while all external particles are massless. Although it is easy to see that setting all $m_i$ equal implies $x_{i,i+1}^2=p_i^2=0$ by~\eqref{dihedral_condition}, it is less obvious that this has no effect on the formula in (\ref{dihedral_all_mass_box_r_formula}). The latter fact can be explained by noticing that these two limits are conformally equivalent (even though the physical interpretation of the two cases is quite different). Using internal masses to regulate the infrared divergences of one- and higher-loop integrals is an old idea; thus, what is interesting here is the simplicity of the case where the internal masses are taken to be finite.

%================================================================================================================
\vspace{-0pt}\subsection{Regge Symmetry}\label{subsec:regge_symmetry}\vspace{-0pt}
%================================================================================================================

Having leveraged known expressions for the volume of geodesic tetrahedra to provide explicit formulas for the all-mass box in all (four-dimensional) space-time signatures, we close this section by highlighting one aspect of this correspondence that we have not made use of. Hyperbolic tetrahedra have a non-obvious Regge symmetry that resembles an identity obeyed by \(6 j\) symbols. Namely, if we treat the lengths of the six sides of the tetrahedron as if they were angular momentum variables and put them into \(6 j\) symbol notation, we have
\begin{equation}
  \begin{Bmatrix}
    l_{12} & l_{23} & l_{13} \\
    l_{34} & l_{14} & l_{24}
  \end{Bmatrix}
\end{equation}
where the first row corresponds to a face of the tetrahedron while columns correspond to opposite sides. This $6j$ symbol obeys a Regge symmetry
\begin{equation}
  \begin{Bmatrix}
    l_{12} & l_{23} & l_{13} \\
    l_{34} & l_{14} & l_{24}
  \end{Bmatrix} \to
    \begin{Bmatrix}
    s - l_{12} & s - l_{23} & l_{13} \\
    s - l_{34} & s - l_{14} & l_{24}
  \end{Bmatrix}
\end{equation}
for \(s = (l_{12} + l_{23} + l_{34} + l_{14}) / 2\). The all-mass box also respects this symmetry, in which four of its side lengths are replaced. Curiously, the volume of the tetrahedron in flat space, given by the Cayley-Menger determinant formula 
\begin{equation}
  \text{vol}(l_{ij})^2 = \frac 1 {2^3 (3!)^2}
  \begin{vmatrix}
    0 & l_{12}^2 & l_{13}^2 & l_{14}^2 & 1 \\
    l_{12}^2 & 0 & l_{23}^2 & l_{24}^2 & 1 \\
    l_{13}^2 & l_{23}^2 & 0 & l_{34}^2 & 1 \\
    l_{14}^2 & l_{24}^2 & l_{34}^2 & 0 & 1 \\
    1 & 1 & 1 & 1 & 0
  \end{vmatrix},
\end{equation}
has the same symmetry~\cite{MR2342290}, as can easily be seen by making the same length substitutions. It would be interesting to understand the physical implications of this discrete symmetry, but we leave this to future work.

%================================================================================================================
\vspace{-0pt}\section{Odd \texorpdfstring{$n$}{n}-gon Integrals in Higher Dimensions}\label{sec:odd_n_integrals}\vspace{-0pt}
%================================================================================================================

In this section we show that $\hat{I}_{n}$ can be computed for odd $n$ using a generalized Gauss-Bonnet theorem, which relates the corresponding $(n{-}1)$-dimensional hyperbolic volume to sums of lower-dimensional volumes (see for example the introduction of~\cite{MR1649192}). The volume of the relevant $(n{-}1)$-dimensional simplices were considered in~\cite{MR1338325}; in particular, this reference showed that the recursion formula we review below satisfies the Schl\"afli differential equations. 
 
The volumes of four- and higher-(even-)dimensional simplices were briefly treated in~\cite{Schnetz:2010pd}. Therein we find the following formula for the Euler characteristic of a hyperbolic \(({n{-}1})\)-dimensional simplex \(\Delta_{n-1}\):
\begin{equation}
  \label{eq:gauss_bonnet}
  \chi(\Delta_{n-1}) = \sum_{j = 0, 2, \dotsc}^{n-1} \frac {2 (-1)^{\frac j 2}}{\operatorname{vol}{(S^j)} \operatorname{vol}(S^{n - j - 2})} \sum_{\sigma \in \text{\(j\)-faces}} \operatorname{vol}(\sigma) \operatorname{polyh}(\sigma),
\end{equation}
where $n$ is assumed to be odd and 
\begin{equation} \label{eq:n_dim_sphere_volume}
\operatorname{vol}(S^k) = \frac {2 \pi^{\frac {k + 1} 2}}{\Gamma(\frac {k + 1} 2)}
\end{equation}
is the volume of the \(k\)-dimensional unit sphere. Since $\Delta_{n-1}$ is a hyperbolic simplex, the volume of each of its faces $\operatorname{vol}(\sigma)$ will also be hyperbolic. Conversely, the polyhedral angles \(\operatorname{polyh}(\sigma)\) can be understood as spherical volumes, as follows.  Consider all the codimension-one faces of the simplex \(\Delta_{n-1}\). Each of these faces is characterized by a normal (or dual) vector, defined in analogy to equation~\eqref{eq:box_dual_vectors}. Any collection of these dual vectors, normalized to unity, determine a spherical simplex---that is, a simplex in signature $(n,0)$. The polyhedral angle of a face $\sigma$ is just the simplicial volume generated by the dual vectors associated with the codimension-one faces of \(\Delta_{n-1}\) that are incident with $\sigma$ (or, more specifically, that contain $\sigma$ as a face).

In order to apply the version of the Gauss-Bonnet formula in eq.~\eqref{eq:gauss_bonnet}, we make use of the fact that $\operatorname{vol}(S^{-1}) = 1$, $\operatorname{vol}(S^0) = 2$, and $\operatorname{polyh}(\Delta_{n-1}) = 1$ by definition.\footnote{For $k<{n{-}1}$, there will be $n{-}k{-}1$ codimension-one faces of $\Delta_{n-1}$ incident with one of its $k$-dimensional faces. Thus, the definition $\operatorname{polyh}(\Delta_{n-1}) = 1$ loosely corresponds to thinking of none of the codimension-one faces as being incident with $\Delta_{n-1}$; more precisely, it follows from defining \(\operatorname{polyh}(\sigma)\) to be the angle subtended by the dual of the cone generated by $\sigma$ (which is equivalent to the definition we offer in the text for $k<n{-}1$)~\cite{Schnetz:2010pd}. The fact that a zero-dimensional sphere has volume 2 follows from defining the volume of a single point to be 1.} For odd $n$, we also have that \(\chi(\Delta_{n-1}) = 1\). We next turn to two explicit examples, to see how~\eqref{eq:gauss_bonnet} works in practice.

%================================================================================================================
\vspace{-0pt}\subsection{The Hyperbolic Triangle Revisited}\label{subsec:odd_n_triangle}\vspace{-0pt}
%================================================================================================================

For a triangle in two dimensions (see also~\cite{Schnetz:2010pd}), the Gauss-Bonnet identity yields
\begin{align}
  1 = \chi(\Delta_2) &= \frac 2 {\operatorname{vol}(S^0) \operatorname{vol}(S^1)} \sum_{\sigma_0 \in \text{\hspace{-.07cm} 0-faces}} \operatorname{vol}(\sigma_0) \operatorname{polyh}(\sigma_0) \nonumber \\
   &\hspace{1cm} + \frac {-2}{\operatorname{vol}(S^2) \operatorname{vol}(S^{-1})} \sum_{\sigma_2 \in \text{\hspace{-.07cm} 2-faces}} \operatorname{vol}(\sigma_2) \operatorname{polyh}(\sigma_2) \, .
\end{align}
Since there is only a single $2$-face (the triangle itself) we can solve for its volume. Using the fact that $\operatorname{polyh}(\Delta_2) = 1$ and plugging in the values~\eqref{eq:n_dim_sphere_volume}, we find
\begin{align}
 \operatorname{vol}(\Delta_2)  &=  \sum_{\sigma_0 \in \text{\hspace{-.07cm} 0-faces}} \operatorname{polyh}(\sigma_0) -  2 \pi  \, .
\end{align}
If we denote the dihedral angles between the edges of this triangle by $\alpha$, $\beta$, and $\gamma$, the corresponding polyhedral angles are \(\pi - \alpha\), \(\pi - \beta\) and \(\pi - \beta\). Thus, we have that
\begin{equation} 
  \operatorname{vol}(\Delta_2) = \pi - \alpha - \beta - \gamma \, ,
\end{equation}
as expected (matching (\ref{eq:hyperbolic_triangle_area})).

%================================================================================================================
\vspace{-0pt}\subsection{The All-Mass Pentagon Integral in Five Dimensions}\label{subsec:odd_n_pentagon}\vspace{-0pt}
%================================================================================================================

Consider now a pentagon in four dimensions, $\Delta_4$. Here equation~\eqref{eq:gauss_bonnet} gives us 
\begin{align}
  1 &= \frac {2}{\operatorname{vol}(S^0) \operatorname{vol}(S^3)} \sum_{\sigma_0 \in \text{\hspace{-.07cm} 0-faces}} \operatorname{vol}(\sigma_0) \operatorname{polyh}(\sigma_0) \\
   &\hspace{1cm} + \frac {-2}{\operatorname{vol}(S^2) \operatorname{vol}(S^1)} \sum_{\sigma_2 \in \text{\hspace{-.07cm} 2-faces}} \operatorname{vol}(\sigma_2) \operatorname{polyh}(\sigma_2) + \frac {2 \operatorname{vol}(\Delta_4) }{\operatorname{vol}(S^4) \operatorname{vol}(S^{-1})} \, , \nonumber
\end{align}
which, upon plugging in the sphere volumes and solving for the volume of the pentagon, becomes
\begin{equation} \label{eq:hyperbolic_four_simplex_vol}
  \operatorname{vol}(\Delta_4) = \frac {4 \pi^2} 3
  - \frac 2 3 \sum_{\sigma_0 \in \text{\hspace{-.07cm} 0-faces}} \operatorname{polyh}(\sigma_0)
  + \frac 1 3 \sum_{\sigma_2 \in \text{\hspace{-.07cm} 2-faces}} \operatorname{vol}(\sigma_2) \operatorname{polyh}(\sigma_2).
\end{equation}
The angles $\operatorname{polyh}(\sigma_0)$ correspond to spherical tetrahedra formed out of four of the vectors dual to the vertices of $\Delta_4$, and similarly each angle $\operatorname{polyh}(\sigma_2)$ corresponds to the angle between a pair of these dual vectors. The volumes $\operatorname{vol}(\sigma_2)$ correspond to hyperbolic triangles formed directly out of the vertices of $\Delta_4$.

We now consider the hyperbolic pentagon whose volume gives $\hat{I}_5$. The kinematic region corresponding to $(4,1)$ signature can be worked out in the same way as for the box---we require that all choices of four of the vertices form a hyperbolic tetrahedron (namely, that they satisfy the constraints given in section~\ref{subsec:box}), and further that the product of all five eigenvalues of $\gram$ is negative, $\det \gram < 0$.

In order to make use of~\eqref{eq:hyperbolic_four_simplex_vol}, we compute the matrix $\gram^*$ as we did for the box, using equation~\eqref{gstar_defined}. These dual vectors are normal to the codimension-one faces of the pentagon, and have unit length. To compute the polyhedral angle of one of the pentagon's vertices $h_i$ in terms of the entries of this matrix, we consider the four codimension-one faces incident with $h_i$---that is, the four tetrahedra formed by the vertices $\{h_i,h_j,h_k,h_l\}$, for any choice of $j,k,l \in \{1,2,3,4,5\} \backslash \{ i \}$. The dual vector normal to each of these faces is labeled by the single vertex it is not incident with; for instance, \(h_i^*\) is normal to the only tetrahedron face not incident with $h_i$. To compute the angle $\operatorname{polyh}(\sigma_{\{h_i\}})$, we therefore compute the spherical tetrahedron formed by the four dual vectors $\{h_j^*,h_k^*,h_l^*,h_m^*\}$ where $j,k,l,m \in \{1,2,3,4,5\} \backslash \{ i \}$. The geometry of this tetrahedron is described by the angles $\cos \theta_{jk} = \gram^*_{jk}$. Thus, we can compute this volume using equation~\eqref{eq:spherical_tetrahedron_vol} after deleting the $i^{\text{th}}$ row and column of $\gram^*$. That is, 
\begin{equation}
\operatorname{polyh}(\sigma_{\{h_i\}}) = V_4\left(\gram^*_{(i)}\right)\, ,
\end{equation}
where $\gram^*_{(i)}$ denotes the $4 \times 4$ matrix that remains after deleting column and row $i$ from $\gram^*$. 

We also need to compute the polyhedral angle of each of the two-dimensional faces of the pentagon. These faces are hyperbolic triangles formed by triples of vertices $\{h_i,h_j,h_k\}$, and are incident with only two of the pentagon's codimension-one faces. The spherical volume formed by the pair of dual vectors normal to these codimension-one faces is therefore
\begin{equation} \label{eq:triangle_dual_cone}
\operatorname{polyh}(\sigma_{\{h_i,h_j,h_k\}}) = \theta_{l,m} = \arccos \big(\gram^*_{l m}\big) \, ,
\end{equation}
namely the angle between $h_l^*$ and $h_m^*$, where $h_l,h_m \notin \{h_i,h_j,h_k\}$.

The final ingredients we need to make use of are just the volumes of the two-dimensional faces themselves, which we know from section~\ref{subsec:triangle}. More precisely, the volume of the face formed by the vertices $\{h_i,h_j,h_k\}$ is given by $\hat{I}_3\big(\gram_{(l m)} \big)$, where again $h_l, h_m \notin \{h_i,h_j,h_k\}$ and the subscript in parentheses denotes deleting these rows and columns.  

Putting this all together, we obtain 
\begin{equation} \label{eq:all_mass_pentagon}
   \hat{I}_5(\gram) = \frac {4 \pi^2}{3}
   - \frac 2 3 \sum_{i = 1}^5 V_4 \big(\gram^*_{(i)} \big)
   + \frac 1 3 \sum_{1 \leq i < j \leq 5} \arccos \big(\gram^*_{ij}\big) \hat{I}_3\big(\gram_{(i j)}\big) \, .
\end{equation}
This gives the Feynman integral $\hat{I}_5$ in terms of lower-dimensional simplicial volumes. Like the all-mass box integral in~\eqref{murakami_yano_formula_for_box}, permutation symmetry is manifest, and the expression involves only classical polylogarithms (although converting the trigonometric functions to logs introduces imaginary arguments). While this integral depends on the solution to five quadratic equations, these equations are individually no more complicated than what was seen in the case of the box.

A similar formula can be derived for the volume of a spherical pentagon. Here the factor of $(-1)^{\frac j 2}$ is absent from equation~\eqref{eq:gauss_bonnet}, and the volumes of the pentagon's faces will also be spherical. The spherical pentagon is thus given by
\begin{equation} \label{eq:spherical_pentagon}
   V_5(\gram) = \frac {4 \pi^2}{3}
   - \frac 2 3 \sum_{i = 1}^5 V_4\big(\gram^*_{(i)}\big)
   - \frac 1 3 \sum_{1 \leq i < j \leq 5} \arccos \big(\gram^*_{i j} \big) V_3 \big(\gram_{(i j)} \big) \quad \text{(mod $\frac 8 3 \pi^2$)},
\end{equation}
where the matrix of dual vectors $\gram^*$ is calculated in the same way as in the hyperbolic case, and subscripts in parentheses again denote deleting these rows and columns. The volume of a spherical triangle $V_3$ was given in~\eqref{eq:spherical_triangle}, and the volume of a spherical tetrahedron $V_4$ was given in~\eqref{eq:spherical_tetrahedron_vol}. Just like for the box, it can be easily checked that $\hat{I}_5(\gram) = V_5(\gram)$ in this region.

We have checked these formulas in a number of ways.  A simple test is to take the simplices to be small.  Then, the effect of the curvature is small and the volume can be approximated by the volume of the simplex in Euclidean space.  We have also checked that the spherical volume~\eqref{eq:spherical_pentagon} constructed out of all right angles evaluates to the appropriate fraction of the embedding sphere (\(\frac 1 4\) for a circle, \(\frac 1 8\) for a two-sphere, \(\frac 1 {16}\) for a three-sphere, etc.).

%================================================================================================================
\vspace{-0pt}\subsection{The Pentagon with Massless Internal Propagators}\label{subsec:odd_n_pentagon_massless}\vspace{-0pt}
%================================================================================================================

Formula~\eqref{eq:all_mass_pentagon} also simplifies when some (or all) of the internal propagators become massless. Let us describe what happens when we take $m_5 \to 0$, which corresponds to sending the vertex \(h_5\) to the boundary of hyperbolic space. To compute the volume of the pentagonal simplex in this limit, we compute the solid angle on the unit three-sphere that this simplex subtends at \(h_5\).  This solid angle is determined by the dual vectors $h_1^*$, $h_2^*$, $h_3^*$, and $h_4^*$ that are normal to the faces of the pentagon incident with $h_5$. 

From~\eqref{gram_matrix} we have \(- \langle h_i, h_5\rangle \to \infty\) for \(i \neq 5\). To see what happens, we rewrite the matrix $\gram$ in a way that separates out index $5$, namely 
\begin{equation}
  \gram =
  \begin{pmatrix}
    \gram_{(5)} & M \\
   M^T & 1
   \end{pmatrix}, \label{eq:break_up_G}
\end{equation}
where $M_i = -\langle h_i, h_5 \rangle$ should be thought of as a column vector of length four. The inverse of $\gram$, which describes the set of dual vectors \(h_i^*\) via~\eqref{gstar_defined}, is then
\begin{equation}
  \gram^{-1} = (1 - M^T \gram_{(5)}^{-1} M)^{-1}
  \begin{pmatrix}
    \gram_{(5)}^{-1} (1 - M^T \gram_{(5)}^{-1} M) + \gram_{(5)}^{-1} M M^T \gram_{(5)}^{-1} &\ \ -\gram_{(5)}^{-1} M \\
    -M^T \gram_{(5)}^{-1} & \ \ 1
  \end{pmatrix}. \label{inverse_gram_m5_separate}
\end{equation}
As the dual vectors \(h_1^*\), \(h_2^*\), \(h_3^*\), and \(h_4^*\) (which all have positive norm) can be individually rescaled by a positive number without affecting the solid angle at $h_5$, we ignore the difference between $\gram^{-1}$ and $\gram^*$ in what follows, and read the Gram matrix of these dual vectors directly off of the top-left \(4 \times 4\) block of~\eqref{inverse_gram_m5_separate}. In the limit \(m_5 \to 0\) (where \(M_i \to \infty\) for \(i = 1, \dots, 4\)), this $4 \times 4$ block becomes
\begin{equation}
  \label{eq:break_up_G_dual}
  \gram_{(5)}^{-1} - \frac{\gram_{(5)}^{-1} M M^T \gram_{(5)}^{-1}}{M^T \gram_{(5)}^{-1} M}.
\end{equation}
This Gram matrix is singular since it has a right eigenvector \(M\) with zero eigenvalue.  Hence, the normal vectors \(\{h_1^*, h_2^*, h_3^*, h_4^*\}\) are linearly dependent.  In fact, since this Gram matrix is computed with a positive-definite scalar product, we have \(\sum_{i = 1}^4 M_i h_i^* = 0\).

It is slightly tricky to define the solid angle generated by a set of linearly dependent vectors.  It may happen that one of these vectors lies inside the cone generated by the others, in which case it does not contribute to the solid angle.  However, this does not happen here; from the positivity conditions on the elements of the Gram matrix for \(h_i\) we know that all the components of the vector \(M_i\) have the same sign.  This fact, together with the relation \(\sum_{i = 1}^4 M_i h_i^* = 0\) implies that none of the vectors \(h_i^*\) lies in the cone generated by the others.

This implies that the spherical simplex defined by $h_1^*$, $h_2^*$, $h_3^*$, and $h_4^*$ spans the full hemisphere bounded by the equatorial sphere to which they all belong. Stated differently, these vectors span half of the volume of a three-dimensional sphere in four dimensions. Thus, we have
\begin{equation}
  \operatorname{polyh}(\{h_5\})\Big|_{m_5 \to \infty} = \frac 1 2 \left(2 \pi^2 \right) = \pi^2.
\end{equation}
The other volumes can all be calculated as before, using the \(- \langle h_i, h_5\rangle \to \infty\) limits of~\eqref{eq:break_up_G} and~\eqref{eq:break_up_G_dual}.

When all the masses are taken to zero, the simplex corresponding to $\hat{I}_5$ is ideal, and all the angles $\operatorname{polyh}(\sigma_{\{h_i\}})$ become $\pi^2$. In this limit, the two-dimensional faces also become ideal triangles, and we have $V_3(\gram_{(ij)}) = \pi$. Taking both of these simplifications into account, \eqref{eq:all_mass_pentagon} becomes 
\begin{equation}
  \hat{I}_5\left(\gram\big|_{m_i = 0}\right) = -2 \pi^2
  + \frac {\pi} 3 \sum_{1 \leq i < j \leq 5}\arccos \big(\gram^*_{ij}\big).
\end{equation}
This can be compared to~\cite{Nandan:2013ip}, where this formula was worked out using different methods (see also~\cite{Schnetz:2010pd}).

%================================================================================================================
\vspace{-0pt}\subsection{All-Mass Integrals in Higher Dimensions}\label{subsec:higher_dim}\vspace{-0pt}
%================================================================================================================

The computational strategy described above generalizes to all odd $n$. In particular,~\eqref{eq:gauss_bonnet} can be recast as
\begin{align}
\hat{I}_n(\gram) &= (-1)^{\frac {n-1} 2} \frac {\pi^{\frac {n} 2}}{\Gamma\left(\frac {n} 2 \right)} - \Bigg( \sum_{j = 0, 2, \dotsc}^{n-3} (-1)^{\frac{n+j-1}{2}} \frac{\Gamma \left(\frac{n-j-1}{2}\right) \Gamma\left(\frac{j+1}{2}\right)}{2\hspace{.04cm} \Gamma\left(\frac{n}{2}\right)} \label{all_mass_higher_odd_dim} \\
&\hspace{4.4cm} \times \hspace{-.4cm} \sum_{i_1 < \dots < i_{n-j-1}} \hspace{-.4cm} \hat{I}_j\big(\gram_{(i_1 \cdots i_{n-j-1})}\big) V_{n-j-1}\big(\gram^*_{i_1 \cdots i_{n-j-1}}\big) \Bigg)\nonumber
\end{align}
after plugging in the volume of the $k$-dimensional unit spheres, $\hat{I}_k$ for all the hyperbolic volumes, and $V_k$ for all the spherical volumes. As can be seen from the first term, this formula is expected to lead to an expression with transcendental weight $\frac{n-1}{2}$. Note that~\eqref{all_mass_higher_odd_dim} is not quite a recursion formula, since it requires computing increasingly higher-dimensional spherical volumes in addition to the lower-point hyperbolic volumes $\hat{I}_{n-2k}$. 

For the $n$-gon with massless internal propagators, the spherical volumes \(V_k\) are all ideal and can be computed as in the case of the massless pentagon; namely, \(V_k\) is given by half the volume of the unit \((k{-}1)\)-sphere. Thus, the Gauss-Bonnet theorem in spherical signature can be used to compute the (internally) massless limit of $\hat{I}_n$ for arbitrarily large (odd) $n$.

%================================================================================================================
%================================================================================================================
\newpage\vspace{-0pt}\section[The Schl\"afli Formula and Branch Cut Structure]{The Schl\"afli Formula and Branch Cut Structure}\label{sec:schlafli}\vspace{-0pt}
%================================================================================================================

Amplitudes only develop branch cuts at kinematic loci where internal propagators go on-shell. It therefore seems worth exploring the interplay of these physical restrictions with the geometry of simplicial volumes. 
A natural tool for doing this is the Schl\"afli formula, which expresses the differential volume of a hyperbolic simplex as a function of the dihedral angles and volumes formed at the intersections of its codimension-one faces~\cite{Schlaefli:1860}: 
\begin{equation}
  \label{eq:schlafli-formula}
  d \text{vol}(\Delta_{n-1}) = -\frac 1 {n - 2} \sum_{\sigma \in (n-3)\text{-faces}} \text{vol}(\sigma) d \theta(\sigma).
\end{equation}
Here $\Delta_{n-1}$ is an $(n{-}1)$-dimensional hyperbolic simplex, the sum is over all codimension-two faces (which are in one-to-one correspondence with intersections of codimension-one faces), and $\theta(\sigma)$ is the dihedral angle formed by $\Delta_{n-1}$ along the face $\sigma$. A similar formula (with opposite sign) holds for spherical  simplices.

Each of the faces of $\Delta_{n-1}$ is itself a simplicial volume, so the Schl\"afli formula can be applied recursively. In particular,~\eqref{eq:schlafli-formula} can be used to determine the symbol~\cite{Goncharov:2010jf} of these volumes, where the letters appearing in the symbol will be just the exponentiated dihedral angles \(\exp (i \theta)\)~\cite{MR1649192,Spradlin:2011wp,Arkani-Hamed:2017ahv,Herrmann:2019upk}. Thus, the Schl\"afli formula must encode the location of all physical branch cuts that appear in $\hat{I}_n$. 

\subsection{Symbols for All \texorpdfstring{$n$}{n}}

When $n$ is even, recursive application of~\eqref{eq:schlafli-formula} to a simplex $\Delta_{n-1}$ will eventually terminate in a sum over its one-dimensional faces. These faces are just the geodesics between pairs of vertices $\{h_i, h_j\}$, namely the bubble integrals considered in section~\ref{subsec:bubble}. It therefore follows from equation~\eqref{eq:bubble_integral} that the first entries of $\hat{I}_n$ will always be drawn from the set of variables $\{r_{ij}\}$ defined in~\eqref{eq:r_vars}. This corresponds to a massive version of the first entry condition considered in~\cite{Gaiotto:2011dt}, similar to what was observed in~\cite{Abreu:2015zaa,Abreu:2017mtm}. 

The second entries will be determined by the dihedral angles formed between pairs of two-dimensional faces. These angles are given by the matrices $\gram^*$ that describe tetrahedra formed by any four vertices of $\Delta_{n-1}$, as per equation~\eqref{gstar_defined}. In particular, specializing to the tetrahedron formed by vertices $\{h_i,h_j,h_k,h_l\}$, the dihedral angle formed along the edge connecting vertices $h_i$ and $h_j$ is given by $\arccos \gram^*_{kl}$, where $\{i,j\} \cap \{k,l\} = \emptyset$. This means that the corresponding symbol entry is $\exp(i \arccos \gram^*_{kl}) = r^*_{kl}$, where $r^*_{kl}$ satisfies the relation
\begin{align}
\gram^*_{kl} = \frac{r^*_{kl} + ({r^*_{kl}})^{-1}}{2} \, 
\end{align}
in analogy with equation~\eqref{eq:r_vars}. Note that these are precisely the variables~\eqref{abcdef_defined} that appear in the Murakami-Yano formula. Solving for $r^*_{kl}$, we have
\begin{align}
r^*_{kl} =  \frac{\det \gram_{(k \neq l)} \pm \sqrt{- \det \gram \det \gram_{(kl)} } }{\sqrt{\det \gram_{(k)}} \sqrt{\det \gram_{(l)}}} \, , \label{eq:r_dual_soln}
\end{align}
where $\gram$ is (minus) the Gram matrix describing the vertices $\{h_i,h_j,h_k,h_l\}$ as usual, and $\gram_{(k \neq l)}$ denotes the matrix $\gram$ with column $k$ and row $l$ deleted. 

Applying this argument iteratively, we deduce that the $j^\text{th}$ symbol entries in $\hat{I}_n$ will be drawn from an analogous set of variables---namely, those given by evaluating~\eqref{eq:r_dual_soln} on the Gram matrices that describe hyperbolic simplices formed out of any $2j$ of the $n$ vertices defining $\hat{I}_n$. Specifically, the Schl\"afli formula gives us 
\begin{align}
\mathcal{S}\big(\hat{I}_n\big) = \sum r_{i_1 i_2} \otimes r^{*(i_1 i_2)}_{i_3 i_4} \otimes r^{*(i_1 i_2 i_3 i_4)}_{i_5 i_6} \otimes \dots \otimes r^{*(i_1 \cdots i_{n-2})}_{i_{n-1} i_n}\, ,\qquad  \text{(even $n$)} \label{eq:symbol_even_n} 
\end{align}
where the sum is over the partitions of \(\{1, \dots, n\}\) as a union of disjoint pairs \(\{i_1, i_2\} \bigger{\cup}\) \( \{i_3, i_4\} \bigger{\cup} \dots \bigger{\cup} \{i_{n-1}, i_n\}\), and where $\smash{r^{*(i_1 \cdots i_{2 j})}_{i_{2 j + 1} i_{2 j + 2}}}$ denotes $r_{i_{2 j + 1} i_{2 j + 2}}^*$ as given in~\eqref{eq:r_dual_soln} when the right hand side is evaluated on the $2j \times 2j$ matrix formed by the rows and columns of the full $n\times n$ Gram matrix with indices $\{i_1,\dots,i_{2 j}\}$. It is worth comparing this formula for $\mathcal{S}\big(\hat{I}_n\big)$ with the results of~\cite{Abreu:2017mtm}; in particular, by comparing~\eqref{eq:r_dual_soln} with equation (D.24) of that paper, one can identify $r^{*(i_1 i_2)}_{i_3 i_4}$ with the double cut of the box integral on propagators $i_1$ and $i_2$, $r^{*(i_1 i_2 i_3 i_4)}_{i_5 i_6}$ with the quad cut of a hexagon on propagators $i_1,\ldots,i_4$, and so on. The full symbol of these integrals is assembled in equation (9.24) of that paper, and can be seen to have the exact same structure as~\eqref{eq:symbol_even_n}. These results can also be compared with those of~\cite{Arkani-Hamed:2017ahv}.

For odd $n$, the recursive application of~\eqref{eq:schlafli-formula} to $\Delta_{n-1}$ will terminate in a sum over its two-dimensional faces. We can read off the corresponding first entries from the triangle integral~\eqref{eq:feyn_triangle}, after converting the arctan to a logarithm:
\begin{align}
\hat{I}_3 (\gram) &= i \log \left( \frac{i(1+\gram_{j k} + \gram_{j l} +\gram_{k l}) + \sqrt{\det \gram}}{i(1+\gram_{j k} + \gram_{j l} +\gram_{k l}) - \sqrt{\det \gram}} \right) \equivL i \log ({R}_{j k l}) \, ,
\end{align}
where $-\gram$ is the Gram matrix formed by any three vertices $\{h_j,h_k,h_l\}$ of $\Delta_{n-1}$, and we denote the corresponding symbol letter by $R_{j k l}$. Subsequent letters can be determined in the same way as for even $n$. Thus, we have
\begin{align}
\mathcal{S}(\hat{I}_n) = \sum R_{i_1 i_2 i_3} \otimes r^{*(i_1 i_2 i_3)}_{i_4 i_5} \otimes r^{*(i_1 i_2 i_3 i_4 i_5)}_{i_6 i_7} \otimes \dots \otimes r^{*(i_1\cdots i_{n-2})}_{i_{n-1} i_n}\, ,\quad  \text{(odd $n$)} \label{eq:symbol_odd_n}
\end{align}
where the sum is over all partitions of $\{1,\dots,n\}$ into one triplet plus pairs, and the Gram matrix defining $r^{*(i_1 \cdots i_{2 j - 1})}_{i_{2 j}\, i_{2 j + 1}}$ in~\eqref{eq:r_dual_soln} is understood to be the submatrix of $\gram$ formed by the rows and columns with indices in $\{i_1,\dots,i_{2 j - 1}\}$.

\subsection{Branch Cuts and Iterated Discontinuities}

To interpret this physically, let us briefly analyze the branch cuts that appear in these symbol entries. Recalling the definition of $r_{ij}$ from~\eqref{eq:r_vars}, and solving for $r_{ij}$ in terms of $x_{ij}$, $m_i$, and $m_j$, we find the two solutions
\begin{align} \label{eq:r_to_kinematics}
r^{\pm}_{ij} = \frac{m_i^2+m_j^2 + x_{ij}^2 \pm \sqrt{\Big(x_{ij}^2+(m_i+m_j)^2 \Big) \Big(x_{ij}^2+(m_i-m_j)^2 \Big)}}{2 m_i m_j} \, .
\end{align}
There are two algebraic branch points in \(r^{\pm}_{i j}\) due to the square root, at the threshold $x_{ij}^2 = - (m_i + m_j)^2$ and at the pseudothreshold $x_{ij}^2 = - (m_i - m_j)^2$.  The Riemann surface of \(r_{i j}\) as a function of \(x_{i j}^2\) can be constructed as follows.  The complex plane with a cut between the two algebraic branch points has the topology of a punctured disk with the boundary being the cut, while the puncture corresponds to the point at infinity.  To obtain the Riemann surface we glue this punctured disk to a second punctured disk and obtain a sphere with two punctures.  These two punctures are logarithmic branch points. Branch cuts associated with internal masses have also been studied for the triangle integral in $4-2\epsilon$ dimensions in~\cite{Abreu:2015zaa}, and are a general feature of massive Feynman integrals. 

It is easy to see from~\eqref{eq:r_dual_soln} that there will be additional algebraic branch cuts in the deeper entries of the symbol, giving rise to a complicated analytic structure. To probe the existence of logarithmic branch cuts, though, we merely need to search for kinematic loci where the symbol letter $\smash{r^{*(i_1 \cdots i_{j-2})}_{i_{j-1} i_j}}$ vanishes or becomes infinite.\footnote{In general, one should first make sure to express a symbol in terms of a multiplicatively independent alphabet of symbol letters to ensure that one doesn't encounter spurious branch cuts that cancel between terms (in particular, when symbol letters are algebraic, this can prove to be surprisingly complicated~\cite{Bourjaily:2019igt}). However, the Schl\"afli formula ensures this will not be a problem insofar as each symbol letter in~\eqref{eq:symbol_even_n} occurs with a unique sequence of letters in front of and behind it; thus, it cannot mix with any other letters.} This happens, for instance, when $\gram_{i_{j-1} i_j} \to \pm \infty$. However, there exist additional logarithmic branch cuts that end on loci depending on multiple kinematic invariants (for instance, where one of the denominator factors in~\eqref{eq:r_dual_soln} vanishes).

A similar set of observations can be made when $n$ is odd. The symbol letter $R_{j k l}$ has logarithmic branch points starting at all three of the thresholds $x_{j k}^2 = - (m_j + m_k)^2$, $x_{j l}^2 = - (m_j + m_l)^2$, and $x_{k l}^2 = - (m_k + m_l)^2$. The letters that appear in subsequent symbol entries are analogous to those appearing for even $n$, and have logarithmic branch cuts that in general depend on on multiple kinematic invariants. 

We note, finally, that the logarithmic branch cuts we have identified in the first and second entries allow for double discontinuities that seem to violate the Steinmann relations~\cite{Steinmann,Steinmann2,Cahill:1973qp} (as they are applied, for instance, in~\cite{Bartels:2008ce,Caron-Huot:2016owq,Dixon:2016nkn,Basso:2017jwq,Caron-Huot:2019bsq}). We leave the resolution of this apparent discrepancy to future work.

%================================================================================================================
%    5. Discussion and Conclusions
%================================================================================================================
\newpage\vspace{-0pt}\section{Conclusions and Open Questions}\label{sec:conclusions}\vspace{-0pt}
%================================================================================================================

In this paper, we have further explored the correspondence between one-loop Feynman integrals and simplicial volumes, expanding on previous studies of the geometry of these integrals~\cite{aomoto1977,Aomoto1992,MR1649192,Davydychev:1997wa,Schnetz:2010pd,Mason:2010pg,Spradlin:2011wp,Nandan:2013ip,Davydychev:2017bbl,Arkani-Hamed:2017ahv,Herrmann:2019upk}. We have focused on the class of all-mass $n$-particle integrals in $n$ dimensions, leaving a study of these integrals in general space-time dimension to a forthcoming companion paper~\cite{n_gons_d_dim}. In $n$ dimensions, these integrals respect a dual conformal symmetry, and evaluate to generalized polylogarithms of uniform transcendental weight $\lfloor n/2\rfloor$ (times a kinematic-independent prefactor).

Using this correspondence, we have provided new dilogarithmic expressions for the all-mass box integral in four dimension and the all-mass pentagon integral in five dimensions, and have additionally studied a number of their kinematic degenerations. Unlike existing dilogarithmic formulas for the all-mass box~\cite{actwu:1961, tHooft:1978jhc, Denner:1991qq}, the form given in~\eqref{murakami_yano_formula_for_box} makes manifest the permutation and conformal symmetries of this integral, and only involves a single algebraic root. The expression for the all-mass pentagon given in~\eqref{eq:all_mass_pentagon} shares these properties, except that it involves a five-orbit of algebraic roots. To our knowledge, the latter integral has not previously appeared in the physics literature (although the limit with massless internal lines was computed in~\cite{Nandan:2013ip}). These expressions for the box and pentagon only involve 16 and 80 dilogarithms, respectively; it is worth wondering whether there exists another form of either function that involves fewer terms.

While we have given formulas for the all-mass box in all (four-dimensional) space-time signatures, and for the pentagon in spherical and hyperbolic signatures, it is worth investigating whether these regions can be understood as part of a more unified geometric picture. For instance, geodesics in the projective model can intersect at points outside of hyperbolic space (understood as the interior of the unit ball centered at $(0,\dots,0,1)$ in $\mathbb{E}^{n-1,1}$). This leads to generalized hyperbolic polytopes, where the exterior vertices are truncated by polar hyperplanes with respect to the quadric corresponding to the boundary of hyperbolic space.  Can these truncated polytopes be used to understand the analytic continuation to other signatures? More generally, it would be interesting to initiate a study of simplicial volumes in signatures beyond the spherical and hyperbolic cases, as we are unaware of this being systematically studied. It could also be instructive to better understand why the expression for the all-mass box given in~\eqref{all_mass_box_r_formula} works in all signatures.

The all-mass box integral famously involves a square root, and this feature is generically shared by the higher-point integrals we have considered. Is it possible to find a \((2 n {-} 1)\)-dimensional simplex whose dihedral angles are all rational (in the sense that their trigonometric functions are all rational), and such that the same conditions are satisfied recursively for all \((2 n {-} 3)\)-dimensional faces? If these simplices exist, do they form a (potentially infinite) discrete set, or a continuous family depending on several variables? Does this set have some density properties? Notably, there are known examples of orthoschemes with essential angles \(\frac \pi p\) for various integers \(p\), which it turns out come in correspondence with Coxeter diagrams (see ref.~\cite{MR1091935}).

In section~\ref{subsec:regge_symmetry}, we have highlighted the existence of an additional Regge symmetry respected by hyperbolic tetrahedra, and consequently by the all-mass box in this signature. It would be interesting to investigate whether this symmetry encodes known---or currently unknown---physical principles. In this vein, it is worth mentioning that the tetrahedron integral also has fascinating connections to Turaev-Viro invariants, R-matrices, and integrability. For instance, there has been recent work on computing these integrals using Yangian symmetry~\cite{Loebbert:2019vcj}.

With the help of the Schl\"afli formula, we have additionally presented an explicit formula for the symbols of these integrals for all $n$. Similar results for the symbols of one-loop integrals can be found in~\mbox{\cite{MR1649192,Spradlin:2011wp,Arkani-Hamed:2017ahv,Abreu:2017mtm}}; in particular, direct analogues of equations~\eqref{eq:r_dual_soln} and~\eqref{eq:symbol_even_n} can be found in~\cite{Abreu:2017mtm}, although it is interesting to note that the formulas found there were derived from a different point of view, using the diagrammatic coaction of Feynman integrals. The symbol is a useful tool for studying the discontinuity structure of these integrals, and can be used to bootstrap integrals and amplitudes even at high transcendental weights (see for example~\cite{Caron-Huot:2019vjl}). In some cases, similar techniques can also be applied to higher-loop integrals to derive their symbol, as shown in~\cite{Herrmann:2019upk}. 

It would in particular be valuable to understand the interplay between the simplicial geometry encoding these symbols and the Steinmann relations. It is possible that some kind of geometric principle is at work here similar to the `cluster adjacency' principle that has been observed in planar maximally supersymmetric Yang-Mills theory~\cite{Drummond:2017ssj,Drummond:2018dfd,Golden:2018gtk,Golden:2019kks,Mago:2019waa}, where an extended version of the Steinmann relations have been observed to hold~\cite{Caron-Huot:2018dsv,Caron-Huot:2019bsq}.  It would also be interesting to see if the recent discussion in ref.~\cite{Arkani-Hamed:2019rds} can be extended to the integrals we studied in this paper. More generally, while we have carried out the beginnings of an analysis of the analytic structure of these integrals, a more in-depth study is called for.

In this paper, we have focused entirely in individual Feynman integrals rather than full amplitudes. However, in~\cite{Mason:2010pg} it was shown that one-loop MHV amplitudes in $\mathcal{N}=4$ supersymmetric Yang-Mills theory are given by the volume of three-dimensional polytopes in \(\mathbb{H}_5\) with no boundary.  In particular, this was demonstrated in the case where all of the propagators have the same mass (or in AdS language, when all four vertices lie on the same horosphere through the infinity twistor \cite{Hodges:2010kq}).  It would be interesting to explore whether this observation could be extended to the case of unequal masses.

Finally, while the connection between Feynman integrals and simplicial volumes breaks down beyond one loop, the integration contours appearing in higher-loop integrals have in many cases been observed to correspond to higher-dimensional Calabi-Yau manifolds~\cite{Brown:2009ta,Brown:2010bw,Bourjaily:2017bsb,Bourjaily:2018ycu,Bourjaily:2018yfy,Festi:2018qip,Broedel:2019kmn,Besier:2019hqd,Bourjaily:2019hmc}. Thus, a more general geometric formulation of Feynman integrals may exists at higher loop orders that could be leveraged to compute these integrals efficiently. Such an interpretation would be especially interesting for integrals that appear in scattering amplitudes at all particle multiplicities, such as those found in~\cite{Bourjaily:2015jna,Bourjaily:2019gqu}.

%================================================================================================================
%    Acknowledgements 
%================================================================================================================
%\vspace{\fill}\vspace{-4pt}
\section*{Acknowledgements}%
\vspace{-4pt}
This project started from discussions and explorations with Matthias Wilhelm. We are also grateful for productive discussions with Dennis M\"uller, Ellis Yuan, and Nima Arkani-Hamed. This project has been supported by an ERC Starting Grant \mbox{(No.\ 757978)} and a grant from the Villum Fonden \mbox{(No.\ 15369)} (JLB,AJM,CV), a Carlsberg Postdoctoral Fellowship \mbox{(CF18-0641)} (AJM), and the STFC Consolidated Grant ``Particle Physics at the Higgs Centre'' (EG). EG is grateful to the CERN theory department for their hospitality while a scientific associate. Finally, we are grateful for the hospitality of the Harvard Center for Mathematical Sciences and Applications and also the Aspen Center for Physics, which is supported by the National Science Foundation under grant \mbox{PHY-1607611}.

\appendix

%================================================================================================================
%    Appendix A:  
%================================================================================================================
\vspace{-0pt}\section[A Short Introduction to the Embedding Formalism]{Short Introduction to the Embedding Formalism}\label{appendix:embedding_formalism}\vspace{-0pt}

%2,3,4,5,6,7
The conformality of $I_n(\gram)$ discussed here is most easily seen in the embedding formalism for inverse propagators. This formalism appears to be used more often than it is explained (see~\cite{Dirac:1936fq,Mack:1969rr,Ferrara:1973yt,Ferrara:1974qk} for early references, and e.g.\ \cite{Weinberg:2010fx,Hodges:2010kq,SimmonsDuffin:2012uy,Abreu:2017ptx,Bourjaily:2019jrk} for more recent presentations and applications). As such, it is worthwhile to provide a reference for its most important ingredients here. 
 
For each external dual-momentum point $x_i^\mu$ and the internal mass $m_i$ associated with the propagator bounding the region it corresponds to (see Figure~\ref{fig:integral_and_dual}), we associate a higher-dimensional vector
\eq{(x_i^\mu,m_i)\mapsto X_i^M\equivR\!\left(\begin{array}{@{}c@{}}x_i^\mu\\x_i^2+m_i^2\\1\end{array}\right)\!\in\mathbb{P}^{n+1}.\label{embedding_formalism_map}}
Similarly, to each loop momentum we associate
\eq{X_{\ell}^M\equivR\!\left(\begin{array}{@{}c@{}}x_\ell^\mu\\x_\ell^2\\1\end{array}\right)\!\in\mathbb{P}^{n+1}\,.\label{embedding_formalism_map_2}}
On the space of $X_i$'s we define an inner product using space-time $x_i^\mu$'s metric $\eta^{\mu\nu}$ according to
\eq{\x{i}{j}\equivR X_i\!\cdot\!X_j\equivR h_{MN}X_i^MX_j^N\quad\text{with}\quad h^{MN}\equivR\left(\begin{array}{@{}c@{$\;$}|@{$\;$}c@{$\;\;$}c@{$\,$}}\\[-16pt]-2\eta^{\mu\nu}&0&0\\[-2pt]
\hline0&0&1\\[-2pt]
0&1&0\\[-2pt]\end{array}\right)\,.\label{embedding_space_metric}}
Using this metric, it is easy to see that  
\eq{\x{i}{j}=x_{ij}^2+m_i^2+m_j^2\,,\qquad\x{\ell}{i}=x_{\ell\,i}^2+m_i^2\,.}
Thus, the propagators appearing in the original integral (\ref{dual_momentum_space_definition}) are now rendered linear in this embedding space. The integration measure over $x_{\ell}$ changes slightly to reflect the embedding map (\ref{embedding_formalism_map}), resulting in
\eq{I_n^0=2\int\!\!\!\proj{d^{n+1}\!X_{\ell}}\delta\big(\x{\ell}{\ell}\big)\frac{1}{\x{\ell}{1}\x{\ell}{2}\x{\ell}{3}\cdots\x{\ell}{n}}\,,\label{embedding_representation_initial_numerator}}
where we have used the notation introduced in~\eqref{projective_notation}.

Because every factor in the denominator of (\ref{embedding_representation_initial_numerator}) is linear in $X_{\ell}$, it is easy to see that there should be {\it two} leading singularities as claimed above---the duplication arising from the quadratic constraint $\delta\big(\x{\ell}{\ell}\big)$ on the final degree of freedom. Moreover, it makes it much easier to Feynman parameterize. As every factor is linear, introducing Feynman parameters is as easy as adding them linearly into 
\eq{\embd{Y}\equivR\sum_i\alpha_i\embd{X}_i\,,}
in terms of which we have
\eq{I_n^0=2\Gamma(n)\int\limits_0^\infty\!\!\!\proj{d^{n-1}\!\vec{\alpha}}\int\!\!\!\proj{d^{n+1}\!X_\ell}\frac{\delta\big(\x{\ell}{\ell}\big)}{\x{\ell}{Y}^n}=\pi^{n/2}\Gamma(n/2)\int\limits_0^\infty\!\!\!\proj{d^{n-1}\!\vec{\alpha}}\frac{1}{\big[\frac{1}{2}\x{Y}{Y}\big]^{\frac{n}{2}}}\,.}
In this form, the conformal symmetry discussed above is made manifest. Namely, if the space-time signature for $\eta^{\mu\nu}$ is $(p,q)$, then the embedding space metric $h^{MN}$ has signature $(p+1,q+1)$; from this, it is easy to see that $I_n$ enjoys an $SO(p+1,q+1)$ symmetry---the conformal group of $\mathbb{R}^{p,q}$. 

%================================================================================================================
%    References (& /Document)
%================================================================================================================
\vspace{20pt}
\newpage
%\bibliographystyle{physics}
%\bibliography{all-massive}

\begin{thebibliography}{10}

\bibitem{n_gons_d_dim}
J.~Bourjaily, E.~Gardi, A.~J. McLeod, and C.~Vergu, ``{All-Mass $n$-gon
  Integrals in $d$ Dimensions}.'' In prep.

\bibitem{Bern:1992em}
Z.~Bern, L.~J. Dixon, and D.~A. Kosower, ``{Dimensionally Regulated One Loop
  Integrals},'' \href{http://dx.doi.org/10.1016/0370-2693(93)90469-X,
  10.1016/0370-2693(93)90400-C}{{\em Phys. Lett.} {\bf B302} (1993)  299--308},
  \href{http://arxiv.org/abs/hep-ph/9212308}{{ arXiv:hep-ph/9212308 [hep-ph]}}.
[Erratum: Phys. Lett.B318,649(1993)].
%%CITATION = HEP-PH/9212308;%%.

\bibitem{Tarasov:1996br}
O.~V. Tarasov, ``{Connection between Feynman Integrals having Different Values
  of the Space-Time Dimension},''
  \href{http://dx.doi.org/10.1103/PhysRevD.54.6479}{{\em Phys. Rev.} {\bf D54}
  (1996)  6479--6490},
\href{http://arxiv.org/abs/hep-th/9606018}{{ arXiv:hep-th/9606018 [hep-th]}}.
%%CITATION = HEP-TH/9606018;%%.

\bibitem{Lee:2009dh}
R.~N. Lee, ``{Space-Time Dimensionality $D$ as Complex Variable: Calculating
  Loop Integrals using Dimensional Recurrence Relation and Analytical
  Properties with Respect to $D$},''
  \href{http://dx.doi.org/10.1016/j.nuclphysb.2009.12.025}{{\em Nucl. Phys.}
  {\bf B830} (2010)  474--492},
\href{http://arxiv.org/abs/0911.0252}{{ arXiv:0911.0252 [hep-ph]}}.
%%CITATION = ARXIV:0911.0252;%%.

\bibitem{Davydychev:1997wa}
A.~I. Davydychev and R.~Delbourgo, ``{A Geometrical Angle on Feynman
  Integrals},'' \href{http://dx.doi.org/10.1063/1.532513}{{\em J. Math. Phys.}
  {\bf 39} (1998)  4299--4334},
\href{http://arxiv.org/abs/hep-th/9709216}{{ arXiv:hep-th/9709216 [hep-th]}}.
%%CITATION = HEP-TH/9709216;%%.

\bibitem{Mason:2010pg}
L.~Mason and D.~Skinner, ``{Amplitudes at Weak Coupling as Polytopes in
  $AdS_{5}$},'' \href{http://dx.doi.org/10.1088/1751-8113/44/13/135401}{{\em J.
  Phys.} {\bf A44} (2011)  135401},
\href{http://arxiv.org/abs/1004.3498}{{ arXiv:1004.3498 [hep-th]}}.
%%CITATION = ARXIV:1004.3498;%%.

\bibitem{actwu:1961}
A.~C.-T. Wu, ``{On the Analytic Properties of the 4-Point Function in
  Perturbation Theory},'' {\em Matematisk-fysiske Meddelelser udgivet af Det
  Kongelige Dnaske Videnskabernes Selskab} {\bf 33} (1961) no. 3, 1--88.
  \href{http://gymarkiv.sdu.dk/MFM/kdvs/mfm%2030-39/mfm-33-3.pdf}{PhD Thesis}.

\bibitem{tHooft:1978jhc}
G.~'t~Hooft and M.~J.~G. Veltman, ``{Scalar One Loop Integrals},''
\href{http://dx.doi.org/10.1016/0550-3213(79)90605-9}{{\em Nucl. Phys.} {\bf
  B153} (1979)  365--401}.
%%CITATION = NUPHA,B153,365;%%.

\bibitem{Denner:1991qq}
A.~Denner, U.~Nierste, and R.~Scharf, ``{A Compact Expression for the Scalar
  One-Loop Four-Point Function},''
\href{http://dx.doi.org/10.1016/0550-3213(91)90011-L}{{\em Nucl. Phys.} {\bf
  B367} (1991)  637--656}.
%%CITATION = NUPHA,B367,637;%%.

\bibitem{Hodges:2010kq}
A.~Hodges, ``{The Box Integrals in Momentum-Twistor Geometry},''
  \href{http://dx.doi.org/10.1007/JHEP08(2013)051}{{\em JHEP} {\bf 1308} (2013)
   051},
\href{http://arxiv.org/abs/1004.3323}{{ arXiv:1004.3323 [hep-th]}}.
%%CITATION = ARXIV:1004.3323;%%.

\bibitem{Nandan:2013ip}
D.~Nandan, M.~F. Paulos, M.~Spradlin, and A.~Volovich, ``{Star Integrals,
  Convolutions and Simplices},''
  \href{http://dx.doi.org/10.1007/JHEP05(2013)105}{{\em JHEP} {\bf 05} (2013)
  105},
\href{http://arxiv.org/abs/1301.2500}{{ arXiv:1301.2500 [hep-th]}}.
%%CITATION = ARXIV:1301.2500;%%.

\bibitem{aomoto1977}
K.~Aomoto, ``{Analytic Structure of {S}chl\"{a}fli Function},''
  \href{http://dx.doi.org/http://projecteuclid.org/euclid.nmj/1118796538}{{\em
  Nagoya Math J.} {\bf 68} (1977)  1--16}.

\bibitem{MR1239859}
E.~B. Vinberg, ``{Volumes of non-{E}uclidean Polyhedra},''
  \href{http://dx.doi.org/10.1070/RM1993v048n02ABEH001011}{{\em Uspekhi Mat.
  Nauk} {\bf 48} (1993) no. 2(290), 17--46}.

\bibitem{MR1338325}
G.~J. Heckman, ``{The Volume of Hyperbolic {C}oxeter Polytopes of Even
  Dimension},'' \href{http://dx.doi.org/10.1016/0019-3577(95)91242-N}{{\em
  Indag. Math. (N.S.)} {\bf 6} (1995) no. 2, 189--196}.

\bibitem{MR1649192}
A.~Goncharov, ``{Volumes of Hyperbolic Manifolds and Mixed {T}ate Motives},''
  \href{http://dx.doi.org/10.1090/S0894-0347-99-00293-3}{{\em J. Amer. Math.
  Soc.} {\bf 12} (1999) no. 2, 569--618}.

\bibitem{Ellis:2007qk}
R.~K. Ellis and G.~Zanderighi, ``{Scalar One-Loop Integrals for QCD},''
  \href{http://dx.doi.org/10.1088/1126-6708/2008/02/002}{{\em JHEP} {\bf 02}
  (2008)  002},
\href{http://arxiv.org/abs/0712.1851}{{ arXiv:0712.1851 [hep-ph]}}.
%%CITATION = ARXIV:0712.1851;%%.

\bibitem{Dixon:2011ng}
L.~J. Dixon, J.~M. Drummond, and J.~M. Henn, ``{The One-Loop Six-Dimensional
  Hexagon Integral and its Relation to MHV Amplitudes in $\mathcal{N}\!=\!4$
  SYM},'' \href{http://dx.doi.org/10.1007/JHEP06(2011)100}{{\em JHEP} {\bf 06}
  (2011)  100},
\href{http://arxiv.org/abs/1104.2787}{{ arXiv:1104.2787 [hep-th]}}.
%%CITATION = ARXIV:1104.2787;%%.

\bibitem{DelDuca:2011ne}
V.~Del~Duca, C.~Duhr, and V.~A. Smirnov, ``{The Massless Hexagon Integral in
  $D\!=\!6$ Dimensions},''
  \href{http://dx.doi.org/10.1016/j.physletb.2011.07.079}{{\em Phys. Lett.}
  {\bf B703} (2011)  363--365},
\href{http://arxiv.org/abs/1104.2781}{{ arXiv:1104.2781 [hep-th]}}.
%%CITATION = ARXIV:1104.2781;%%.

\bibitem{DelDuca:2011jm}
V.~Del~Duca, C.~Duhr, and V.~A. Smirnov, ``{The One-Loop One-Mass Hexagon
  Integral in $D\!=\!6$ Dimensions},''
  \href{http://dx.doi.org/10.1007/JHEP07(2011)064}{{\em JHEP} {\bf 07} (2011)
  064},
\href{http://arxiv.org/abs/1105.1333}{{ arXiv:1105.1333 [hep-th]}}.
%%CITATION = ARXIV:1105.1333;%%.

\bibitem{DelDuca:2011wh}
V.~Del~Duca, L.~J. Dixon, J.~M. Drummond, C.~Duhr, J.~M. Henn, and V.~A.
  Smirnov, ``{The One-Loop Six-Dimensional Hexagon Integral with Three Massive
  Corners},'' \href{http://dx.doi.org/10.1103/PhysRevD.84.045017}{{\em Phys.
  Rev.} {\bf D84} (2011)  045017},
\href{http://arxiv.org/abs/1105.2011}{{ arXiv:1105.2011 [hep-th]}}.
%%CITATION = ARXIV:1105.2011;%%.

\bibitem{Papadopoulos:2014lla}
C.~G. Papadopoulos, ``{Simplified Differential Equations Approach for Master
  Integrals},'' \href{http://dx.doi.org/10.1007/JHEP07(2014)088}{{\em JHEP}
  {\bf 07} (2014)  088},
\href{http://arxiv.org/abs/1401.6057}{{ arXiv:1401.6057 [hep-ph]}}.
%%CITATION = ARXIV:1401.6057;%%.

\bibitem{Spradlin:2011wp}
M.~Spradlin and A.~Volovich, ``{Symbols of One-Loop Integrals From Mixed Tate
  Motives},'' \href{http://dx.doi.org/10.1007/JHEP11(2011)084}{{\em JHEP} {\bf
  11} (2011)  084},
\href{http://arxiv.org/abs/1105.2024}{{ arXiv:1105.2024 [hep-th]}}.
%%CITATION = ARXIV:1105.2024;%%.

\bibitem{Kozlov:2015kol}
M.~G. Kozlov and R.~N. Lee, ``{One-Loop Pentagon Integral in $d$ Dimensions
  from Differential Equations in $\epsilon$-Form},''
  \href{http://dx.doi.org/10.1007/JHEP02(2016)021}{{\em JHEP} {\bf 02} (2016)
  021},
\href{http://arxiv.org/abs/1512.01165}{{ arXiv:1512.01165 [hep-ph]}}.
%%CITATION = ARXIV:1512.01165;%%.

\bibitem{Aomoto1992}
K.~Aomoto, ``{Hyperlogarithmic Expansion and the Volume of a Hyperbolic
  Simplex},'' {\em Banach Center Publications} {\bf 27} (1992) no. 1, 9--21.
  \url{http://eudml.org/doc/262870}.

\bibitem{Schnetz:2010pd}
O.~Schnetz, ``{The Geometry of One-Loop Amplitudes},''
\href{http://arxiv.org/abs/1010.5334}{{ arXiv:1010.5334 [hep-th]}}.
%%CITATION = ARXIV:1010.5334;%%.

\bibitem{Davydychev:2017bbl}
A.~I. Davydychev, ``{Four-Point Function in General Kinematics through
  Geometrical Splitting and Reduction},''
  \href{http://dx.doi.org/10.1088/1742-6596/1085/5/052016}{{\em J. Phys. Conf.
  Ser.} {\bf 1085} (2018) no. 5, 052016},
\href{http://arxiv.org/abs/1711.07351}{{ arXiv:1711.07351 [hep-th]}}.
%%CITATION = ARXIV:1711.07351;%%.

\bibitem{Arkani-Hamed:2017ahv}
N.~Arkani-Hamed and E.~Y. Yuan, ``{One-Loop Integrals from Spherical
  Projections of Planes and Quadrics},''
\href{http://arxiv.org/abs/1712.09991}{{ arXiv:1712.09991 [hep-th]}}.
%%CITATION = ARXIV:1712.09991;%%.

\bibitem{Herrmann:2019upk}
E.~Herrmann and J.~Parra-Martinez, ``{Logarithmic Forms and Differential
  Equations for Feynman Integrals},''
\href{http://arxiv.org/abs/1909.04777}{{ arXiv:1909.04777 [hep-th]}}.
%%CITATION = ARXIV:1909.04777;%%.

\bibitem{Loebbert:2019vcj}
F.~Loebbert, D.~M{\"u}ller, and H.~M{\"u}nkler, ``{Yangian Bootstrap for
  Conformal Feynman Integrals},''
\href{http://arxiv.org/abs/1912.05561}{{ arXiv:1912.05561 [hep-th]}}.
%%CITATION = ARXIV:1912.05561;%%.

\bibitem{MR2154824}
J.~Murakami and M.~Yano, ``{On the Volume of a Hyperbolic and Spherical
  Tetrahedron},'' {\em Comm. Anal. Geom.} {\bf 13} (2005) no. 2, 379--400.

\bibitem{MR2917101}
J.~Murakami, ``{Volume Formulas for a Spherical Tetrahedron},''
  \href{http://dx.doi.org/10.1090/S0002-9939-2012-11182-7}{{\em Proc. Amer.
  Math. Soc.} {\bf 140} (2012) no. 9, 3289--3295}.

\bibitem{Goncharov:2010jf}
A.~B. Goncharov, M.~Spradlin, C.~Vergu, and A.~Volovich, ``{Classical
  Polylogarithms for Amplitudes and Wilson Loops},''
  \href{http://dx.doi.org/10.1103/PhysRevLett.105.151605}{{\em Phys. Rev.
  Lett.} {\bf 105} (2010)  151605},
\href{http://arxiv.org/abs/1006.5703}{{ arXiv:1006.5703 [hep-th]}}.
%%CITATION = ARXIV:1006.5703;%%.

\bibitem{Schlaefli:1860}
L.~Schl{\"a}fli, ``{On the Multiple Integral $\int^n\!dx\, dy \cdots dz$ Whose
  Limits are $p_1=a_1 x + b_1 y + \cdots + h_1 z \geq 0, \,
  p_2\geq0,\cdots,p_n\geq0$ and $x^2 + y^2 + \cdots + z^2<1$},'' {\em
  Quart.~J.~of Math.} {\bf 3} (1858-1860)  54--68,97--108.

\bibitem{Abreu:2017mtm}
S.~Abreu, R.~Britto, C.~Duhr, and E.~Gardi, ``{Diagrammatic Hopf Algebra of Cut
  Feynman Integrals: the One-Loop Case},''
  \href{http://dx.doi.org/10.1007/JHEP12(2017)090}{{\em JHEP} {\bf 12} (2017)
  090},
\href{http://arxiv.org/abs/1704.07931}{{ arXiv:1704.07931 [hep-th]}}.
%%CITATION = ARXIV:1704.07931;%%.

\bibitem{Bourjaily:2018aeq}
J.~L. Bourjaily, A.~J. McLeod, M.~von Hippel, and M.~Wilhelm, ``{Rationalizing
  Loop Integration},'' \href{http://dx.doi.org/10.1007/JHEP08(2018)184}{{\em
  JHEP} {\bf 08} (2018)  184},
\href{http://arxiv.org/abs/1805.10281}{{ arXiv:1805.10281 [hep-th]}}.
%%CITATION = ARXIV:1805.10281;%%.

\bibitem{ArkaniHamed:2010gh}
N.~Arkani-Hamed, J.~L. Bourjaily, F.~Cachazo, and J.~Trnka, ``{Local Integrals
  for Planar Scattering Amplitudes},''
  \href{http://dx.doi.org/10.1007/JHEP06(2012)125}{{\em JHEP} {\bf 1206} (2012)
   125},
\href{http://arxiv.org/abs/1012.6032}{{ arXiv:1012.6032 [hep-th]}}.
%%CITATION = ARXIV:1012.6032;%%.

\bibitem{gram}
{MacTutor History of Mathematics Archive}.
\newblock \url{http://mathshistory.st-andrews.ac.uk/Biographies/Gram.html}.

\bibitem{Alday:2009dv}
L.~F. Alday, D.~Gaiotto, and J.~Maldacena, ``{Thermodynamic Bubble Ansatz},''
  \href{http://dx.doi.org/10.1007/JHEP09(2011)032}{{\em JHEP} {\bf 09} (2011)
  032},
\href{http://arxiv.org/abs/0911.4708}{{ arXiv:0911.4708 [hep-th]}}.
%%CITATION = ARXIV:0911.4708;%%.

\bibitem{Golden:2013xva}
J.~Golden, A.~B. Goncharov, M.~Spradlin, C.~Vergu, and A.~Volovich, ``{Motivic
  Amplitudes and Cluster Coordinates},''
  \href{http://dx.doi.org/10.1007/JHEP01(2014)091}{{\em JHEP} {\bf 1401} (2014)
   091},
\href{http://arxiv.org/abs/1305.1617}{{ arXiv:1305.1617 [hep-th]}}.
%%CITATION = ARXIV:1305.1617;%%.

\bibitem{Dixon:2013eka}
L.~J. Dixon, J.~M. Drummond, M.~von Hippel, and J.~Pennington, ``{Hexagon
  Functions and the Three-Loop Remainder Function},''
  \href{http://dx.doi.org/10.1007/JHEP12(2013)049}{{\em JHEP} {\bf 1312} (2013)
   049},
\href{http://arxiv.org/abs/1308.2276}{{ arXiv:1308.2276 [hep-th]}}.
%%CITATION = ARXIV:1308.2276;%%.

\bibitem{Dixon:2016apl}
L.~J. Dixon, M.~von Hippel, A.~J. McLeod, and J.~Trnka, ``{Multi-Loop
  Positivity of the Planar $\mathcal{N}\!=\!4$ SYM Six-Point Amplitude},''
  \href{http://dx.doi.org/10.1007/JHEP02(2017)112}{{\em JHEP} {\bf 02} (2017)
  112},
\href{http://arxiv.org/abs/1611.08325}{{ arXiv:1611.08325 [hep-th]}}.
%%CITATION = ARXIV:1611.08325;%%.

\bibitem{Drummond:2006rz}
J.~Drummond, J.~Henn, V.~Smirnov, and E.~Sokatchev, ``{Magic Identities for
  Conformal Four-Point Integrals},''
  \href{http://dx.doi.org/10.1088/1126-6708/2007/01/064}{{\em JHEP} {\bf 0701}
  (2007)  064},
\href{http://arxiv.org/abs/hep-th/0607160}{{ arXiv:hep-th/0607160}}.
%%CITATION = HEP-TH/0607160;%%.

\bibitem{Bern:2006ew}
Z.~Bern, M.~Czakon, L.~J. Dixon, D.~A. Kosower, and V.~A. Smirnov, ``{The
  Four-Loop Planar Amplitude and Cusp Anomalous Dimension in Maximally
  Supersymmetric Yang-Mills Theory},''
  \href{http://dx.doi.org/10.1103/PhysRevD.75.085010}{{\em Phys. Rev.} {\bf
  D75} (2007)  085010},
\href{http://arxiv.org/abs/hep-th/0610248}{{ arXiv:hep-th/0610248 [hep-th]}}.
%%CITATION = HEP-TH/0610248;%%.

\bibitem{Bern:2007ct}
Z.~Bern, J.~Carrasco, H.~Johansson, and D.~Kosower, ``{Maximally Supersymmetric
  Planar Yang-Mills Amplitudes at Five Loops},''
  \href{http://dx.doi.org/10.1103/PhysRevD.76.125020}{{\em Phys. Rev.} {\bf
  D76} (2007)  125020},
\href{http://arxiv.org/abs/0705.1864}{{ arXiv:0705.1864 [hep-th]}}.
%%CITATION = ARXIV:0705.1864;%%.

\bibitem{Alday:2007hr}
L.~F. Alday and J.~M. Maldacena, ``{Gluon Scattering Amplitudes at Strong
  Coupling},'' \href{http://dx.doi.org/10.1088/1126-6708/2007/06/064}{{\em
  JHEP} {\bf 06} (2007)  064},
\href{http://arxiv.org/abs/0705.0303}{{ arXiv:0705.0303 [hep-th]}}.
%%CITATION = ARXIV:0705.0303;%%.

\bibitem{Bern:2008ap}
Z.~Bern, L.~J. Dixon, D.~A. Kosower, R.~Roiban, M.~Spradlin, C.~Vergu, and
  A.~Volovich, ``{The Two-Loop Six-Gluon MHV Amplitude in Maximally
  Supersymmetric Yang-Mills Theory},''
  \href{http://dx.doi.org/10.1103/PhysRevD.78.045007}{{\em Phys. Rev.} {\bf
  D78} (2008)  045007},
\href{http://arxiv.org/abs/0803.1465}{{ arXiv:0803.1465 [hep-th]}}.
%%CITATION = ARXIV:0803.1465;%%.

\bibitem{Drummond:2008vq}
J.~Drummond, J.~Henn, G.~Korchemsky, and E.~Sokatchev, ``{Dual Superconformal
  Symmetry of Scattering Amplitudes in $\mathcal{N}\!=\!4$ super Yang-Mills
  Theory},'' \href{http://dx.doi.org/10.1016/j.nuclphysb.2009.11.022}{{\em
  Nucl. Phys.} {\bf B828} (2010)  317--374},
\href{http://arxiv.org/abs/0807.1095}{{ arXiv:0807.1095 [hep-th]}}.
%%CITATION = ARXIV:0807.1095;%%.

\bibitem{doi:10.1063/1.523697}
B.~G. Nickel, ``{Evaluation of Simple Feynman Graphs},''
  \href{http://dx.doi.org/10.1063/1.523697}{{\em Journal of Mathematical
  Physics} {\bf 19} (1978) no. 3, 542--548}.

\bibitem{2019arXiv190801141R}
D.~{Rudenko}, ``{Rational Elliptic Surfaces and the Trigonometry of
  Tetrahedra},'' \href{http://arxiv.org/abs/1908.01141}{{ arXiv:1908.01141
  [math.AG]}}.

\bibitem{Drummond:2007au}
J.~M. Drummond, J.~Henn, G.~P. Korchemsky, and E.~Sokatchev, ``{Conformal Ward
  Identities for Wilson Loops and a Test of the Duality with Gluon
  Amplitudes},'' \href{http://dx.doi.org/10.1016/j.nuclphysb.2009.10.013}{{\em
  Nucl. Phys.} {\bf B826} (2010)  337--364},
\href{http://arxiv.org/abs/0712.1223}{{ arXiv:0712.1223 [hep-th]}}.
%%CITATION = 0712.1223;%%.

\bibitem{Drummond:2007aua}
J.~M. Drummond, G.~P. Korchemsky, and E.~Sokatchev, ``{Conformal Properties of
  Four-Gluon Planar Amplitudes and Wilson loops},''
  \href{http://dx.doi.org/10.1016/j.nuclphysb.2007.11.041}{{\em Nucl. Phys.}
  {\bf B795} (2008)  385--408},
\href{http://arxiv.org/abs/0707.0243}{{ arXiv:0707.0243 [hep-th]}}.
%%CITATION = 0707.0243;%%.

\bibitem{Alday:2009yn}
L.~F. Alday and J.~Maldacena, ``{Null Polygonal Wilson Loops and Minimal
  Surfaces in Anti- de-Sitter Space},''
  \href{http://dx.doi.org/10.1088/1126-6708/2009/11/082}{{\em JHEP} {\bf 11}
  (2009)  082},
\href{http://arxiv.org/abs/0904.0663}{{ arXiv:0904.0663 [hep-th]}}.
%%CITATION = 0904.0663;%%.

\bibitem{Bourjaily:2013mma}
J.~L. Bourjaily, S.~Caron-Huot, and J.~Trnka, ``{Dual-Conformal Regularization
  of Infrared Loop Divergences and the {\it Chiral} Box Expansion},''
  \href{http://dx.doi.org/10.1007/JHEP01(2015)001}{{\em JHEP} {\bf 1501} (2015)
   001},
\href{http://arxiv.org/abs/1303.4734}{{ arXiv:1303.4734 [hep-th]}}.
%%CITATION = ARXIV:1303.4734;%%.

\bibitem{Alday:2009zm}
L.~F. Alday, J.~M. Henn, J.~Plefka, and T.~Schuster, ``{Scattering into the
  Fifth Dimension of $\mathcal{N}\!=\!4$ super Yang-Mills},''
  \href{http://dx.doi.org/10.1007/JHEP01(2010)077}{{\em JHEP} {\bf 1001} (2010)
   077},
\href{http://arxiv.org/abs/0908.0684}{{ arXiv:0908.0684 [hep-th]}}.
%%CITATION = ARXIV:0908.0684;%%.

\bibitem{Henn:2010bk}
J.~M. Henn, S.~G. Naculich, H.~J. Schnitzer, and M.~Spradlin,
  ``{Higgs-Regularized Three-Loop Four-Gluon Amplitude in $\mathcal{N}\!=\!4$
  SYM: Exponentiation and Regge Limits},''
  \href{http://dx.doi.org/10.1007/JHEP04(2010)038}{{\em JHEP} {\bf 04} (2010)
  038},
\href{http://arxiv.org/abs/1001.1358}{{ arXiv:1001.1358 [hep-th]}}.
%%CITATION = 1001.1358;%%.

\bibitem{MR2342290}
P.~P. Boalch, ``{Regge and {O}kamoto Symmetries},''
  \href{http://dx.doi.org/10.1007/s00220-007-0328-x}{{\em Comm. Math. Phys.}
  {\bf 276} (2007) no. 1, 117--130}.

\bibitem{Gaiotto:2011dt}
D.~Gaiotto, J.~Maldacena, A.~Sever, and P.~Vieira, ``{Pulling the Straps of
  Polygons},'' \href{http://dx.doi.org/10.1007/JHEP12(2011)011}{{\em JHEP} {\bf
  12} (2011)  011},
\href{http://arxiv.org/abs/1102.0062}{{ arXiv:1102.0062 [hep-th]}}.
%%CITATION = ARXIV:1102.0062;%%.

\bibitem{Abreu:2015zaa}
S.~Abreu, R.~Britto, and H.~Gr{\"o}nqvist, ``{Cuts and Coproducts of Massive
  Triangle Diagrams},'' \href{http://dx.doi.org/10.1007/JHEP07(2015)111}{{\em
  JHEP} {\bf 07} (2015)  111},
\href{http://arxiv.org/abs/1504.00206}{{ arXiv:1504.00206 [hep-th]}}.
%%CITATION = ARXIV:1504.00206;%%.

\bibitem{Bourjaily:2019igt}
J.~L. Bourjaily, A.~J. McLeod, C.~Vergu, M.~Volk, M.~Von~Hippel, and
  M.~Wilhelm, ``{Rooting Out Letters: Octagonal Symbol Alphabets and Algebraic
  Number Theory},''
\href{http://arxiv.org/abs/1910.14224}{{ arXiv:1910.14224 [hep-th]}}.
%%CITATION = ARXIV:1910.14224;%%.

\bibitem{Steinmann}
O.~Steinmann, ``{\"Uber den Zusammenhang Zwischen den Wightmanfunktionen und
  der Retardierten Kommutatoren},''
  \href{http://dx.doi.org/10.3929/ethz-a-000107369}{{\em Helv. Physica Acta}
  {\bf 33} (1960)  257}.

\bibitem{Steinmann2}
O.~Steinmann, ``{Wightman-Funktionen und Retardierte Kommutatoren. II},'' {\em
  Helv. Physica Acta} {\bf 33} (1960)  347.
  \url{www.e-periodica.ch/cntmng?pid=hpa-001:1960:33::1079}.

\bibitem{Cahill:1973qp}
K.~E. Cahill and H.~P. Stapp, ``{Optical Theorems and Steinmann Relations},''
\href{http://dx.doi.org/10.1016/0003-4916(75)90006-8}{{\em Annals Phys.} {\bf
  90} (1975)  438}.
%%CITATION = APNYA,90,438;%%.

\bibitem{Bartels:2008ce}
J.~Bartels, L.~N. Lipatov, and A.~Sabio~Vera, ``{BFKL Pomeron, Reggeized Gluons
  and Bern-Dixon-Smirnov Amplitudes},''
  \href{http://dx.doi.org/10.1103/PhysRevD.80.045002}{{\em Phys. Rev.} {\bf
  D80} (2009)  045002},
\href{http://arxiv.org/abs/0802.2065}{{ arXiv:0802.2065 [hep-th]}}.
%%CITATION = ARXIV:0802.2065;%%.

\bibitem{Caron-Huot:2016owq}
S.~Caron-Huot, L.~J. Dixon, A.~McLeod, and M.~von Hippel, ``{Bootstrapping a
  Five-Loop Amplitude Using Steinmann Relations},''
  \href{http://dx.doi.org/10.1103/PhysRevLett.117.241601}{{\em Phys. Rev.
  Lett.} {\bf 117} (2016) no. 24, 241601},
\href{http://arxiv.org/abs/1609.00669}{{ arXiv:1609.00669 [hep-th]}}.
%%CITATION = ARXIV:1609.00669;%%.

\bibitem{Dixon:2016nkn}
L.~J. Dixon, J.~Drummond, T.~Harrington, A.~J. McLeod, G.~Papathanasiou, and
  M.~Spradlin, ``{Heptagons from the Steinmann Cluster Bootstrap},''
  \href{http://dx.doi.org/10.1007/JHEP02(2017)137}{{\em JHEP} {\bf 02} (2017)
  137},
\href{http://arxiv.org/abs/1612.08976}{{ arXiv:1612.08976 [hep-th]}}.
%%CITATION = ARXIV:1612.08976;%%.

\bibitem{Basso:2017jwq}
B.~Basso and L.~J. Dixon, ``{Gluing Ladder Feynman Diagrams into Fishnets},''
  \href{http://dx.doi.org/10.1103/PhysRevLett.119.071601}{{\em Phys. Rev.
  Lett.} {\bf 119} (2017) no. 7, 071601},
\href{http://arxiv.org/abs/1705.03545}{{ arXiv:1705.03545 [hep-th]}}.
%%CITATION = ARXIV:1705.03545;%%.

\bibitem{Caron-Huot:2019bsq}
S.~Caron-Huot, L.~J. Dixon, F.~Dulat, M.~Von~Hippel, A.~J. McLeod, and
  G.~Papathanasiou, ``{The Cosmic Galois Group and Extended Steinmann Relations
  for Planar $\mathcal{N} = 4$ SYM Amplitudes},''
\href{http://arxiv.org/abs/1906.07116}{{ arXiv:1906.07116 [hep-th]}}.
%%CITATION = ARXIV:1906.07116;%%.

\bibitem{MR1091935}
R.~Kellerhals, ``{On {S}chl\"{a}fli's Reduction Formula},''
  \href{http://dx.doi.org/10.1007/BF02571335}{{\em Math. Z.} {\bf 206} (1991)
  no. 2, 193--210}. \url{https://doi.org/10.1007/BF02571335}.

\bibitem{Caron-Huot:2019vjl}
S.~Caron-Huot, L.~J. Dixon, F.~Dulat, M.~von Hippel, A.~J. McLeod, and
  G.~Papathanasiou, ``{Six-Gluon Amplitudes in Planar $\mathcal{N}\!=\!4$
  super-Yang-Mills Theory at Six and Seven Loops},''
  \href{http://dx.doi.org/10.1007/JHEP08(2019)016}{{\em JHEP} {\bf 08} (2019)
  016},
\href{http://arxiv.org/abs/1903.10890}{{ arXiv:1903.10890 [hep-th]}}.
%%CITATION = ARXIV:1903.10890;%%.

\bibitem{Drummond:2017ssj}
J.~Drummond, J.~Foster, and {\"O}.~G{\"u}rdo\v{g}an, ``{Cluster Adjacency
  Properties of Scattering Amplitudes in $\mathcal{N}=4$ Supersymmetric
  Yang-Mills Theory},''
  \href{http://dx.doi.org/10.1103/PhysRevLett.120.161601}{{\em Phys. Rev.
  Lett.} {\bf 120} (2018) no. 16, 161601},
\href{http://arxiv.org/abs/1710.10953}{{ arXiv:1710.10953 [hep-th]}}.
%%CITATION = ARXIV:1710.10953;%%.

\bibitem{Drummond:2018dfd}
J.~Drummond, J.~Foster, and {\"O}.~G{\"u}rdo\v{g}an, ``{Cluster Adjacency
  Beyond MHV},'' \href{http://dx.doi.org/10.1007/JHEP03(2019)086}{{\em JHEP}
  {\bf 03} (2019)  086},
\href{http://arxiv.org/abs/1810.08149}{{ arXiv:1810.08149 [hep-th]}}.
%%CITATION = ARXIV:1810.08149;%%.

\bibitem{Golden:2018gtk}
J.~Golden and A.~J. Mcleod, ``{Cluster Algebras and the Subalgebra
  Constructibility of the Seven-Particle Remainder Function},''
  \href{http://dx.doi.org/10.1007/JHEP01(2019)017}{{\em JHEP} {\bf 01} (2019)
  017},
\href{http://arxiv.org/abs/1810.12181}{{ arXiv:1810.12181 [hep-th]}}.
%%CITATION = ARXIV:1810.12181;%%.

\bibitem{Golden:2019kks}
J.~Golden, A.~J. McLeod, M.~Spradlin, and A.~Volovich, ``{The Sklyanin Bracket
  and Cluster Adjacency at All Multiplicity},''
  \href{http://dx.doi.org/10.1007/JHEP03(2019)195}{{\em JHEP} {\bf 03} (2019)
  195},
\href{http://arxiv.org/abs/1902.11286}{{ arXiv:1902.11286 [hep-th]}}.
%%CITATION = ARXIV:1902.11286;%%.

\bibitem{Mago:2019waa}
J.~Mago, A.~Schreiber, M.~Spradlin, and A.~Volovich, ``{Yangian Invariants and
  Cluster Adjacency in $\mathcal{N}\!=\!4$ Yang-Mills},''
  \href{http://dx.doi.org/10.1007/JHEP10(2019)099}{{\em JHEP} {\bf 10} (2019)
  099},
\href{http://arxiv.org/abs/1906.10682}{{ arXiv:1906.10682 [hep-th]}}.
%%CITATION = ARXIV:1906.10682;%%.

\bibitem{Caron-Huot:2018dsv}
S.~Caron-Huot, L.~J. Dixon, M.~von Hippel, A.~J. McLeod, and G.~Papathanasiou,
  ``{The Double Pentaladder Integral to All Orders},''
  \href{http://dx.doi.org/10.1007/JHEP07(2018)170}{{\em JHEP} {\bf 07} (2018)
  170},
\href{http://arxiv.org/abs/1806.01361}{{ arXiv:1806.01361 [hep-th]}}.
%%CITATION = ARXIV:1806.01361;%%.

\bibitem{Arkani-Hamed:2019rds}
N.~Arkani-Hamed, T.~Lam, and M.~Spradlin, ``{Non-Perturbative Geometries for
  Planar $\mathcal{N}\!=\!4$ SYM Amplitudes},''
\href{http://arxiv.org/abs/1912.08222}{{ arXiv:1912.08222 [hep-th]}}.
%%CITATION = ARXIV:1912.08222;%%.

\bibitem{Brown:2009ta}
F.~C.~S. Brown, ``{On the Periods of Some Feynman Integrals},''
\href{http://arxiv.org/abs/0910.0114}{{ arXiv:0910.0114 [math.AG]}}.
%%CITATION = ARXIV:0910.0114;%%.

\bibitem{Brown:2010bw}
F.~Brown and O.~Schnetz, ``{A K3 in $\phi^4$},''
\href{http://arxiv.org/abs/1006.4064}{{ arXiv:1006.4064 [math.AG]}}.
%%CITATION = ARXIV:1006.4064;%%.

\bibitem{Bourjaily:2017bsb}
J.~L. Bourjaily, A.~J. McLeod, M.~Spradlin, M.~von Hippel, and M.~Wilhelm,
  ``{Elliptic Double-Box Integrals: Massless Scattering Amplitudes beyond
  Polylogarithms},''
  \href{http://dx.doi.org/10.1103/PhysRevLett.120.121603}{{\em Phys. Rev.
  Lett.} {\bf 120} (2018) no. 12, 121603},
\href{http://arxiv.org/abs/1712.02785}{{ arXiv:1712.02785 [hep-th]}}.
%%CITATION = ARXIV:1712.02785;%%.

\bibitem{Bourjaily:2018ycu}
J.~L. Bourjaily, Y.-H. He, A.~J. Mcleod, M.~Von~Hippel, and M.~Wilhelm,
  ``{Traintracks Through Calabi-Yaus: Amplitudes Beyond Elliptic
  Polylogarithms},''
  \href{http://dx.doi.org/10.1103/PhysRevLett.121.071603}{{\em Phys. Rev.
  Lett.} {\bf 121} (2018) no. 7, 071603},
\href{http://arxiv.org/abs/1805.09326}{{ arXiv:1805.09326 [hep-th]}}.
%%CITATION = ARXIV:1805.09326;%%.

\bibitem{Bourjaily:2018yfy}
J.~L. Bourjaily, A.~J. McLeod, M.~von Hippel, and M.~Wilhelm, ``{A (Bounded)
  Bestiary of Feynman Integral Calabi-Yau Geometries},''
  \href{http://dx.doi.org/10.1103/PhysRevLett.122.031601}{{\em Phys. Rev.
  Lett.} {\bf 122} (2019) no. 3, 031601},
\href{http://arxiv.org/abs/1810.07689}{{ arXiv:1810.07689 [hep-th]}}.
%%CITATION = ARXIV:1810.07689;%%.

\bibitem{Festi:2018qip}
D.~Festi and D.~van Straten, ``{Bhabha Scattering and a Special Pencil of K3
  Surfaces},''
\href{http://arxiv.org/abs/1809.04970}{{ arXiv:1809.04970 [math.AG]}}.
%%CITATION = ARXIV:1809.04970;%%.

\bibitem{Broedel:2019kmn}
J.~Broedel, C.~Duhr, F.~Dulat, R.~Marzucca, B.~Penante, and L.~Tancredi, ``{An
  Analytic Solution for the Equal-Mass Banana Graph},''
\href{http://arxiv.org/abs/1907.03787}{{ arXiv:1907.03787 [hep-th]}}.
%%CITATION = ARXIV:1907.03787;%%.

\bibitem{Besier:2019hqd}
M.~Besier, D.~Festi, M.~Harrison, and B.~Naskrecki, ``{Arithmetic and Geometry
  of a K3 Surface Emerging from Virtual Corrections to Drell-Yan Scattering},''
\href{http://arxiv.org/abs/1908.01079}{{ arXiv:1908.01079 [math.AG]}}.
%%CITATION = ARXIV:1908.01079;%%.

\bibitem{Bourjaily:2019hmc}
J.~L. Bourjaily, A.~J. McLeod, C.~Vergu, M.~Volk, M.~Von~Hippel, and
  M.~Wilhelm, ``{Embedding Feynman Integral (Calabi-Yau) Geometries in Weighted
  Projective Space},''
\href{http://arxiv.org/abs/1910.01534}{{ arXiv:1910.01534 [hep-th]}}.
%%CITATION = ARXIV:1910.01534;%%.

\bibitem{Bourjaily:2015jna}
J.~L. Bourjaily and J.~Trnka, ``{Local Integrand Representations of All
  Two-Loop Amplitudes in Planar SYM},''
  \href{http://dx.doi.org/10.1007/JHEP08(2015)119}{{\em JHEP} {\bf 08} (2015)
  119},
\href{http://arxiv.org/abs/1505.05886}{{ arXiv:1505.05886 [hep-th]}}.
%%CITATION = ARXIV:1505.05886;%%.

\bibitem{Bourjaily:2019gqu}
J.~L. Bourjaily, E.~Herrmann, C.~Langer, A.~J. McLeod, and J.~Trnka,
  ``{All-Multiplicity Non-Planar MHV Amplitudes in sYM at Two Loops},''
\href{http://arxiv.org/abs/1911.09106}{{ arXiv:1911.09106 [hep-th]}}.
%%CITATION = ARXIV:1911.09106;%%.

\bibitem{Dirac:1936fq}
P.~A.~M. Dirac, ``{Wave Equations in Conformal Space},''
\href{http://dx.doi.org/10.2307/1968455}{{\em Annals Math.} {\bf 37} (1936)
  429--442}.
%%CITATION = ANMAA,37,429;%%.

\bibitem{Mack:1969rr}
G.~Mack and A.~Salam, ``{Finite Component Field Representations of the
  Conformal Group},''
\href{http://dx.doi.org/10.1016/0003-4916(69)90278-4}{{\em Annals Phys.} {\bf
  53} (1969)  174--202}.
%%CITATION = APNYA,53,174;%%.

\bibitem{Ferrara:1973yt}
S.~Ferrara, A.~F. Grillo, and R.~Gatto, ``{Tensor Representations of Conformal
  Algebra and Conformally Covariant Operator Product Expansion},''
\href{http://dx.doi.org/10.1016/0003-4916(73)90446-6}{{\em Annals Phys.} {\bf
  76} (1973)  161--188}.
%%CITATION = APNYA,76,161;%%.

\bibitem{Ferrara:1974qk}
S.~Ferrara, ``{Supergauge Transformations on the Six-Dimensional Hypercone},''
\href{http://dx.doi.org/10.1016/0550-3213(74)90305-8}{{\em Nucl. Phys.} {\bf
  B77} (1974)  73--90}.
%%CITATION = NUPHA,B77,73;%%.

\bibitem{Weinberg:2010fx}
S.~Weinberg, ``{Six-Dimensional Methods for Four-Dimensional Conformal Field
  Theories},'' \href{http://dx.doi.org/10.1103/PhysRevD.82.045031}{{\em Phys.
  Rev.} {\bf D82} (2010)  045031},
\href{http://arxiv.org/abs/1006.3480}{{ arXiv:1006.3480 [hep-th]}}.
%%CITATION = ARXIV:1006.3480;%%.

\bibitem{SimmonsDuffin:2012uy}
D.~Simmons-Duffin, ``{Projectors, Shadows, and Conformal Blocks},''
  \href{http://dx.doi.org/10.1007/JHEP04(2014)146}{{\em JHEP} {\bf 04} (2014)
  146},
\href{http://arxiv.org/abs/1204.3894}{{ arXiv:1204.3894 [hep-th]}}.
%%CITATION = ARXIV:1204.3894;%%.

\bibitem{Abreu:2017ptx}
S.~Abreu, R.~Britto, C.~Duhr, and E.~Gardi, ``{Cuts from Residues: the One-Loop
  Case},'' \href{http://dx.doi.org/10.1007/JHEP06(2017)114}{{\em JHEP} {\bf 06}
  (2017)  114},
\href{http://arxiv.org/abs/1702.03163}{{ arXiv:1702.03163 [hep-th]}}.
%%CITATION = ARXIV:1702.03163;%%.

\bibitem{Bourjaily:2019jrk}
J.~L. Bourjaily, F.~Dulat, and E.~Panzer, ``{Manifestly Dual-Conformal Loop
  Integration},''
\href{http://arxiv.org/abs/1901.02887}{{ arXiv:1901.02887 [hep-th]}}.
%%CITATION = ARXIV:1901.02887;%%.

\end{thebibliography}
%\end{document}
\providecommand{\href}[2]{#2}\begingroup\raggedright\endgroup
\end{document}